\documentclass[12pt]{article}
\pdfoutput=1

\usepackage{jheppub}
\usepackage{amsmath,amssymb,euscript,array,mathrsfs,appendix,marvosym,calc,array}
\usepackage{arydshln}
\usepackage{graphicx}
\usepackage[normalem]{ulem}

\usepackage{blkarray}
\usepackage[normalem]{ulem}

\usepackage{comment}
\usepackage{nicefrac}

\usepackage{tikz} 
\usepackage{tikz-feynman}
\usetikzlibrary{decorations.pathmorphing}
\usetikzlibrary{decorations.markings}

\usetikzlibrary{calc}
\usetikzlibrary{arrows}
\usetikzlibrary{trees}
\usetikzlibrary{decorations.pathmorphing}
\usetikzlibrary{decorations.markings}

\usepackage{ctable}
\AtBeginDocument{%
  
}

\newcolumntype{C}[1]{>{\centering\arraybackslash$}m{#1}<{$}}
\newlength{\mycolwd}                                         
\newcommand\blank[1]{#1}
\renewcommand\blank[1]{}
\def\Buildrel#1\over#2\under#3{\mathrel{\mathop{\kern0pt
#2}\limits^{#1}_{#3}}}

\usepackage{color}
\definecolor{lightgray}{gray}{0.75}

\def\AA{{\mathfrak L}}

\def\CF{{\cal F}}

\newcommand{\Tr}{\operatorname{Tr}}

\newcommand{\Ad}{\operatorname{ad}}
\newcommand{\AD}{\operatorname{Ad}}

\def\B0{{\boldsymbol 0}}

\def\det{{\rm det}}

\def\Dbarslash{\,\,{\raise.15ex\hbox{/}\mkern-12mu {\bar D}}}
\def\Dslash{\,\,{\raise.15ex\hbox{/}\mkern-12mu D}}
\def\delslash{\,\,{\raise.15ex\hbox{/}\mkern-9mu \partial}}
\def\delbarslash{\,\,{\raise.15ex\hbox{/}\mkern-9mu {\bar\partial}}}

\newcommand{\EQ}[1]{\begin{equation}\begin{split} #1
\end{split}\end{equation}}

\newcommand{\AL}[1]{\begin{align} #1
\end{align}}
\newcommand{\SP}[1]{\begin{equation}\begin{split} #1
\end{split}\end{equation}}

\title{Integrability of the $\lambda$-deformation of the PCM with spectators}
\author{Riccardo Borsato$^1$, Georgios Itsios$^2$, J. Luis Miramontes$^1$ {\normalfont and} Konstantinos Siampos$^3$}
\affiliation{
$^1$Instituto Gallego de F\'isica de Altas Energ\'ias (IGFAE),\\[2pt]
and Departamento de F\'\i sica de Part\'\i culas,\\[2pt]
Universidad de  Santiago de Compostela\\[4pt]
$^2$Institut f\"{u}r Physik, Humboldt-Universit\"{a}t zu Berlin,\\[2pt]
IRIS Geb\"{a}ude, Zum Gro{\ss}en \\[2pt]
Windkanal 2, 12489 Berlin, Germany\\[4pt]
$^3$Laboratory of Theoretical Physics, School of Physics,\\[2pt]
Aristotle University of Thessaloniki,\\[4pt] 
54124 Thessaloniki, Greece
}
\emailAdd{riccardo.borsato@usc.es}
\emailAdd{georgios.itsios@physik.hu-berlin.de}
\emailAdd{jluis.miramontes@usc.es}
\emailAdd{ksiampos@auth.gr}
\preprint{HU-EP-24/24}

\abstract{We construct a generalisation of the $\lambda$-deformation of the Principal Chiral Model (PCM) where we deform just a subgroup $F$ of the full symmetry group $G$. We find that demanding Lax integrability imposes a crucial restriction, namely that  the coset $F\backslash G$ must be symmetric. Surprisingly, we also find that (when $F$ is non-abelian) integrability requires that  the term in the action involving only the spectator fields should have a specific $\lambda$-dependence, which is a curious modification of the procedure expected from the known $F=G$ case. The resulting Lax connection has a novel analytical structure, with four single poles as opposed to the two poles of the cases of the PCM and of the standard $\lambda$-deformation. We also explicitly work out the example of $G=SU(2)$ and $F=U(1)$, discussing its renormalisation group flow  to two loops.
}

\notoc

\begin{document}

\maketitle

\newpage

\tableofcontents

\section{Introduction}
The so-called ``$\lambda$-deformation'' is an integrable 2-dimensional $\sigma$-model first constructed in~\cite{Sfetsos:2013wia}. It is controlled by the deformation parameter $\lambda$ restricted to taking values in $0< \lambda< 1$. The $\sigma$-model can be understood as an interpolation between two known integrable models: at $\lambda=0$ the Wess--Zumino--Witten (WZW) model on the Lie group $G$, while at $\lambda=1$  the $\sigma$-model obtained by applying non-abelian T-duality (NATD) to the Principal Chiral Model (PCM) on $G$.\footnote{To the best of our knowledge, when the procedure of NATD is applied to the full group $G$, then the Lax integrability was first established in~\cite{Mohammedi:2008vd}, see also~\cite{Sfetsos:2013wia}, while for the case of NATD on $F\subset G$ in~\cite{Borsato:2016pas}.}
For small $\lambda$, the model can be understood as a \emph{relevant}\footnote{In fact, the deforming operator does not satisfy the marginality condition of~\cite{Chaudhuri:1988qb}, which in the case of compact Lie groups is a necessary and sufficient condition for integrably marginal deformations of WZW models.} deformation of WZW driven by $\kappa_{ab}\mathcal{J}_+^a\bar{\mathcal J_-}^b$, where $\kappa_{ab}$ is a bilinear form on the Lie algebra $\mathfrak{g}=Lie(G)$ and $\mathcal{J}_+,\bar{\mathcal J_-}$ are the chiral/antichiral currents of the WZW model.

In recent years, the $\lambda$-deformation has been the subject of numerous studies and generalisations. Although a Lax connection was constructed in~\cite{Sfetsos:2013wia}, the strong integrability of the model was established in~\cite{Georgiou:2019plp}. The renormalisation group of the $\sigma$-model was studied in~\cite{Itsios:2014lca,Sfetsos:2014jfa,Georgiou:2015nka}, see also~\cite{Appadu:2015nfa}. In~\cite{Vicedo:2015pna,Hoare:2015gda,Klimcik:2015gba} the model was shown to be Poisson--Lie dual to the $\eta$-deformation of~\cite{Klimcik:2002zj,Klimcik:2008eq}. The $\lambda$-deformation is, in fact, meant to realise a $q$-deformation of the symmetries of the undeformed model, in the case when $q$ is a root of unity \cite{Hoare:2011nd,Hoare:2011wr,Hollowood:2014rla}.

The construction of the $\lambda$-deformation was generalized to the case of symmetric coset spaces in~\cite{Hollowood:2014rla}, and later to semi-symmetric supercosets in~\cite{Hollowood:2014qma}.  Importantly, in all cases the $\lambda\to 1$ limit is related to the NATD of the $\sigma$-models, and there is no value of $\lambda$ for which the $\lambda$-deformation reduces directly to the PCM, the coset or the supercoset $\sigma$-models. Still, in this paper we follow a terminology commonly used in the literature and, to distinguish our case from the (super)coset constructions, we speak of ``$\lambda$-deformation of the PCM''. This is only meant in the sense that the ``seed'' $\sigma$-model of the construction is the PCM, although there is no value of $\lambda$ for which the PCM action is recovered without dualisation. In fact this kind of terminology makes more sense in the Hamiltonian formalism, where the deformation involves a modification of the Poisson brackets of the PCM~\cite{Rajeev:1988hq,Balog:1993es}. The analogues of giant magnons and the classical spectral curve of the latter were discussed in~\cite{Appadu:2017xku} and~\cite{Hollowood:2019ajq}, respectively.
Other generalisations include an ``asymmetric''version~\cite{Driezen:2019ykp}, and D-branes in $\lambda$-deformations were also considered~\cite{Driezen:2018glg}. 

The $\lambda$-deformation attracted a lot of attention because of the potential applications in string theory and the AdS/CFT correspondence. In particular, various supergravity backgrounds for $\lambda$-deformations were constructed. In some of these cases, the limit $\lambda\to 1$ yields the NATD on the \emph{bosonic} group of isometries of the original maximally supersymmetric backgrounds. This is true for example for the $\lambda$-deformed background of  $AdS_3\times S^3\times M^4$  of~\cite{Sfetsos:2014cea,Itsios:2023kma,Itsios:2023uae}, of the $AdS_5\times S^5$ one of~\cite{Demulder:2015lva} and backgrounds that contain undeformed AdS spaces~\cite{Itsios:2019izt}.
The construction of~\cite{Hollowood:2014qma}, instead,  in the $\lambda\to 1$ limit reduces  to NATD on the full \emph{super}group of isometries, and the corresponding  supergravity backgrounds  were derived in the case of $AdS_2\times S^2\times T^6$~\cite{Borsato:2016zcf}, $AdS_3\times S^3\times T^4$~\cite{Chervonyi:2016ajp} and $AdS_5\times S^5$~\cite{Borsato:2016ose}. The $\lambda$-deformation of the superstring was also analysed in the pure spinor formalism~\cite{Benitez:2019oaw}.

In this paper we want to address a natural and simple question: is it possible to construct a $\lambda$-deformation in the presence of spectators? Obviously, we want to look at this problem in the non-trivial case, namely when the dynamics of the spectator fields mix with those of the fields participating in the $\lambda$-deformation. After constructing this generalisation of the $\lambda$-deformation, we may build new integrable models, possibly with potential applications in the AdS/CFT correspondence. A proposal for this construction with spectators was actually put forward already in~\cite{Sfetsos:2013wia}, but the integrability  was not discussed there. In fact, we will show that in order to have integrability, the construction of~\cite{Sfetsos:2013wia} for the $\lambda$-deformation with spectators must be first modified and then subjected to additional conditions. 

In this paper we will focus on the case of the PCM. In particular, we will  construct a $\lambda$-deformation of the PCM on a Lie group $G$, in such a way that only a subgroup $F\subset G$ is $\lambda$-deformed. The name of ``spectators'', then, is justified by the fact that the degrees of freedom in the coset $F\backslash G$ are minimally affected by the deformation. The expectation, that will be confirmed in the paper, is that in an appropriate $\lambda\to 1$ limit this model reduces to the $\sigma$-model  obtained when applying NATD to the PCM on $G$, but when only $F\subset G$ is dualised.

For the construction we will  need to make a technical assumption requiring that the restriction to $\mathfrak{f}$ (the Lie algebra of $F$) of the bilinear form of $\mathfrak{g}$ (the Lie algebra of $G$) is non-degenerate. This assumption is not necessary to construct the NATD on $F\subset G$, but if we do not assume it the construction of the $\lambda$-deformation with spectators may be  ill defined. We refer to section~\ref{sec:deg} for comments on this point. Importantly, we will show that integrability puts  important constraints on the underlying algebraic structure. In fact, although we can construct the $\lambda$-deformation with spectators for generic choices of $G$ and the subgroup $F$, we find that a  Lax connection can be written down only if the coset $F\backslash G$ is \emph{symmetric}.\footnote{Strictly speaking there is another option, which is when $G$ is of direct-product form, $G=F\times H$. However, this case is  trivial and uninteresting, because it is equivalent to a $\lambda$-model of~\cite{Sfetsos:2013wia} on $F$ and a decoupled PCM on $H$.} This extra condition of having a $\mathbb Z_2$ structure  is reminiscent of the analogous condition found in~\cite{Eichenherr:1979ci} when demanding integrability for coset $\sigma$-models.\footnote{Our construction is not meant to be the only possible one, and our results do no exclude the possibility of similar deformations involving spectator fields that may be integrable even in the absence of a $\mathbb Z_2$ grading. In particular, it would be interesting to explore the possibility of constructing other integrable deformations with spectator fields in terms of asymmetric gauged WZW action (see e.g.~\cite{Driezen:2020job}).}
Another crucial point is that in our construction we need to introduce an additional parameter $\alpha$,  and we show that (when $F$ is non-abelian) integrability fixes $\alpha=\frac{1+\lambda}{2}$. According to our results, the construction of~\cite{Sfetsos:2013wia} should not have a Lax connection for generic $\lambda$, because there $\alpha=1$.

The structure of the paper is as follows. In Section~\ref{sec:action} we give the definition of the model and introduce the first technical ingredients. In Section~\ref{sec:sym} we discuss the symmetries of the $\sigma$-model, and in Section~\ref{sec:lim} we discuss the limits $\lambda\to 0$ and $\lambda\to 1$. In Section~\ref{sec:deg} we comment on the obstacles that one may face when trying to generalize our construction to the case where the restriction to $\mathfrak{f}$ of the bilinear form in $\mathfrak{g}$ is degenerate. Section~\ref{sec:int} contains our main results, because after deriving the equations of motion we show that they can be put in the form of the zero-curvature condition for a Lax connection when certain conditions are met. In Section~\ref{sec:ex} we work out an explicit example taking $G=SU(2)$ and $F=U(1)$, also discussing its RG flow to two loops.
Finally, in section~\ref{sec:concl} we end with some conclusions and future directions.

\section{The action of the deformed model}\label{sec:action}

Our starting point is the action of a PCM\footnote{The conventions for the volume form are $d^2x=d\tau d\sigma$, and $\partial_\pm=\frac{\partial}{\partial \sigma^\pm}$ where $\sigma^\pm=\tau\pm\sigma$.\label{normalization.action}}
\EQ{
S_\text{PCM}[g] &= -\frac{\kappa^2}{\pi}\int d^2 x\, \Tr\left\{g^{-1}\partial_+ g\, g^{-1}\partial_- g\right\},
}
where $g$ takes values in a Lie group $G$. We will consider both compact and non-compact Lie groups, and the
only assumption that we need is that the Lie algebra $\mathfrak{g}$ is equipped with a non-degenerate,
symmetric, and ad-invariant bilinear form. For simplicity, although not necessary, we 
have  in mind a matrix realisation of $\mathfrak{g}$, so that for the bilinear form we  take the one induced
by the trace $\Tr$ of products of matrices. When the matrix realisation corresponds to the adjoint representation, this bilinear form is the Killing form, and the condition of being non-degenerate means that $\mathfrak{g}$ is semi-simple. However, it is important to point out that our construction is not restricted to this case.\footnote{See for example~\cite{Nappi:1993ie} for an application in physics of an example of a non-semisimple algebra admitting a non-degenerate bilinear form.}
The PCM is known to be integrable with Lax connection 
\EQ{
\AA_\pm(z) = \frac{z}{z\pm1}\, J_\pm, \qquad J_\pm= g^{-1}\partial_\pm g,
}
where $z$ is a complex spectral parameter.
We now want to consider a subgroup $F\subset G$ and gauge its left action. The standard way to do it is to introduce  gauge fields $A_\pm \in \mathfrak f$ and the action of the gauged PCM (gPCM) model
\EQ{
S_\text{gPCM}[g,A_\mu]&= -\frac{\kappa^2}{\pi}\int d^2 x\, \Tr\left\{\left(\partial_+ g g^{-1} + A_+\right)\, \left(\partial_- g g^{-1}+ A_- \right)\right\},
\label{eq:gPCM}
}
which is in fact invariant under the gauge transformation
\EQ{
g\to U g,\qquad
 A_\pm \to U\left(\partial_\pm+A_\pm\right) U^{-1},
\qquad U\in F\subset G.
\label{eq:GaugeTranf-gPCM}
}
However, as we will see in section~\ref{sec:int}, the requirement that the $\lambda$-deformation in the presence of spectators is integrable  forces us to build it using a \emph{deformation} of the gauged PCM action written above. In order to write this deformed action, we need to specify one extra assumption about the algebraic structure. In particular, throughout this paper we will assume that not only the bilinear form on $\mathfrak{g}$ is non-degenerate, but also its restriction to the Lie subalgebra $\mathfrak{f}$. Notice that this assumption is not needed when constructing the gPCM model, but it turns out that it leads to important simplifications when constructing the $\lambda$-deformed action. We refer to section~\ref{sec:deg} for comments on the case when the restriction of $\Tr$ to $\mathfrak{f}$ is degenerate. 

As usual, the bilinear form of $\mathfrak{g}$ can be used to identify the dual $\mathfrak{f}^*$ of the Lie algebra $\mathfrak{f}$, and the requirement that the restriction of $\Tr$ to $\mathfrak{f}$ is non-degenerate is equivalent to saying that $\mathfrak{f}^*=\mathfrak{f}$.
Because of the non-degeneracy, we can consider the vector-space decomposition of the full Lie algebra as
\EQ{
\mathfrak{g}=\mathfrak{f}\oplus \mathfrak{f}^\perp,
\label{eq:decomp-f-fperp}
}
where the ``orthogonal part'' $\mathfrak{f}^\perp$ satisfies
\EQ{
\Tr\{\mathfrak{f}\,\mathfrak{f}^\perp\}=0.
}
Notice that the above decomposition as a direct sum is to be understood only from the point of view of \emph{vector spaces}. In particular, we are not assuming that $\mathfrak{f}$ and $\mathfrak{f}^\perp$ commute. Still, it is easy to see that the non-degeneracy and the ad-invariance of the bilinear form, together with the assumption that $\mathfrak{f}$ closes into a Lie algebra ($[\mathfrak{f},\mathfrak{f}]\subset \mathfrak{f}$), put an important restriction; namely
\EQ{
[\mathfrak{f},\mathfrak{f}^\perp]\subset \mathfrak{f}^\perp.
\label{eq:comm-f-fperp}
}
These are the only extra properties that we will use in this section, but in section~\ref{sec:int} we will show that integrability imposes an extra restriction on the algebraic structure. In particular, at this point the commutator $[\mathfrak{f}^\perp,\mathfrak{f}^\perp]$ may take values both in $\mathfrak{f}$ and in $\mathfrak{f}^\perp$. Later, we will show that integrability imposes additional conditions on the algebraic structure, and the most interesting cases arise when $[\mathfrak{f}^\perp,\mathfrak{f}^\perp]$ takes value only in $\mathfrak{f}$. Therefore,  we will restrict to cases when $\mathfrak{g}$ admits a $\mathbb Z_2$ decomposition that is compatible with the Lie bracket, so that $\mathfrak{f}$ is the Lie subalgebra with eigenvalue $1$ under the $\mathbb Z_2$ automorphism, while $\mathfrak{f}^\perp$ is the subspace with eigenvalue $-1$.

We now have all the information needed to write the deformation of the gPCM action that we will use to construct the $\lambda$-deformation. It is
\EQ{
\widetilde S_\text{gPCM}[g,A_\mu;\alpha]&= -\frac{\kappa^2}{\pi}\int d^2 x\, \Tr\left\{\left(\partial_+ g g^{-1} + A_+\right)^\mathfrak{f}\, \left(\partial_- g g^{-1}+ A_- \right)^\mathfrak{f} \right.\\[5pt]
&\hspace{3.5cm}
\left.+ \alpha\left(\partial_+g g^{-1}\right)^\perp \left(\partial_-g g^{-1}\right)^\perp\right\}.
\label{eq:gPCM-modified}
}
Here we are using a shorthand notation, denoting by $X^{\mathfrak{f}}$ and $X^{\perp}$ the projections of a generic element $X\in \mathfrak{g}$ onto $\mathfrak{f}$ and $\mathfrak{f}^\perp$, respectively. We also introduced a positive deformation parameter $\alpha$. It is easy to see that for $\alpha=1$ this action reduces to the standard gPCM action~\eqref{eq:gPCM}. Importantly, the action~\eqref{eq:gPCM-modified} is still invariant under the gauge transformation~\eqref{eq:GaugeTranf-gPCM}, but now the fact that this statement holds also for $\alpha\neq 1$ is only possible thanks to the property~\eqref{eq:comm-f-fperp}.\footnote{In fact, if we define the projector $P^\perp$ onto $\mathfrak{f}^\perp$, one can check that~\eqref{eq:comm-f-fperp} implies $P^\perp \Ad_u=\Ad_u P^\perp$ and $P^\perp \AD_U=\AD_U P^\perp$, for $u\in\mathfrak{f}$ and $U\in F$.}

As in the standard $\lambda$-deformation, we now want to couple this (deformation) of the gPCM to a gauged WZW (gWZW) model. The difference with the standard construction of~\cite{Sfetsos:2013wia} is that now we take the $F/F$ gWZW, as opposed to $G/G$. The action of the gWZW is 
\EQ{
&S_\text{gWZW}^{F/F}[{\cal F},A_\mu]
=S_\text{WZW}[\CF]+
\frac{k}{\pi} \int d^2x\, \Tr\left\{-A_+ \partial_-\CF\CF^{-1}
 +A_- \CF^{-1}\partial_+\CF\right.\\
&\quad\quad\quad\quad\quad\quad\quad\left. +\CF^{-1}A_+\CF A_-  -A_+A_- \right\}\,,
\label{eq:gWZW}}
where the WZW action at level $k$ is given by
\EQ{
&S_\text{WZW}[\CF]=-\frac{k}{2\pi} \int d^2x\, \Tr\left\{\CF^{-1}\partial_+\CF 
\CF^{-1}\partial_-\CF\right\}+S_\text{WZ}[\CF]\,,\\\
&S_\text{WZ}[\CF]=\frac{k}{12\pi} \int d^3 x\, \epsilon^{abc}\Tr\left\{\CF^{-1}\partial_a\CF\, \CF^{-1}\partial_b\CF\, \CF^{-1}\partial_c\CF\right\}\,.
\label{eq:WZW}
}
Here $\CF\in F$, and  $A_\pm \in \mathfrak{f}$ are the same gauge fields that we introduced for the gPCM. The last term in~\eqref{eq:WZW} is the, so-called, Wess--Zumino (WZ) term and $k$ is the level of the gWZW action. The gWZW action is invariant under the gauge transformation
\EQ{
\CF\to U\CF U^{-1},\qquad A_\pm \to U\left(\partial_\pm+A_\pm\right) U^{-1},
\qquad U\in F\subset G.
\label{eq:GaugeTranf-gWZW}
}
Because of the assumption of non-degeneracy of the restriction of $\Tr$ to $\mathfrak{f}$, this gWZW action is well defined~\cite{Nappi:1993ie} and the field $\CF$ actually provides $\text{dim}(F)$ degrees of freedom.
The generalisation to the case of degenerate restriction is problematic precisely because of the gWZW model, see the discussion in section~\ref{sec:deg}.

In analogy with the construction of~\cite{Sfetsos:2013wia}, we are now ready to define the action of the $\lambda$-deformation in the presence of spectators as
\EQ{
S[g,{\cal F},A_\mu;\alpha]=\widetilde S_\text{gPCM}[g,A_\mu;\alpha] + S_\text{gWZW}^{F/F}[{\cal F},A_\mu].
\label{eq:Action}
}
Notice that this model has a total of dim$(G)$ degrees of freedom, because of the gauge invariance
\EQ{
g\to U g,\qquad
\CF\to U\CF U^{-1},\qquad A_\pm \to U\left(\partial_\pm+A_\pm\right) U^{-1},
\qquad U\in F\subset G.
\label{eq:GaugeTranf}
} 
The action is controlled by a total of three parameters: the coupling $\kappa$ of the gPCM, the level $k$ of the gWZW\footnote{We remind that $k$ needs to be quantised in the case of compact groups. When the group is non-compact there is no quantisation condition.} and the parameter $\alpha$. From now on, as  usually done for this kind of deformation, we will parameterise the relative magnitude of the couplings $\kappa$ and $k$ with the parameter $\lambda$, which is defined as
\EQ{
\lambda=\frac{k}{k+\kappa^2}\qquad \Longrightarrow \qquad \kappa^2= \frac{1-\lambda}{\lambda}\, k\,,
\label{eq:Couplings}
}
thus $\lambda$ takes values in $(0,1)$.
 The action~\eqref{eq:Action} appeared also in Section 5 of~\cite{Sfetsos:2013wia}, although there the integrability of the model was not discussed.
In fact, in section~\ref{sec:int} we will show that to ensure 
integrability at generic values of $\lambda$ we cannot take $\alpha=1$ and we need some extra algebraic structure.

\vspace{12pt}

At this point one can notice that in the action~\eqref{eq:Action} the gauge fields $A_\pm$ are not dynamical, and they can be integrated out to obtain an action just for $\CF$ and $g$. In particular, the equations of motion for $A_\pm$ give us the solutions 
\EQ{
&
A_+= \left(\lambda\,  \AD_{\CF}^{-1} -1\right)^{-1} \,\left((1-\lambda) j_+^\mathfrak{f}
- \lambda\, \CF^{-1}\partial_+\CF\right),\\[5pt]
&
A_-= \left(\lambda\,  \AD_{\CF} -1\right)^{-1} \,\left((1-\lambda) j_-^\mathfrak{f}
+ \lambda\, \partial_-\CF \CF^{-1}\right),
\label{eq:EqsGaugeFields}
}
where we defined
\EQ{
j_\pm=\partial_\pm gg^{-1} = j_\pm^{\mathfrak{f}}+j_\pm^\perp,
\label{eq:current}
}
and we are using the notation $\AD_\CF X=\CF X\CF^{-1}$.\footnote{Notice that $\AD_\CF^{-1} X=\CF^{-1} X\CF=\AD_{\CF^{-1}}X$, where $\AD_\CF$ is an orthogonal matrix.\label{footnote.Adjoint}}
Then, the final form of the action~\eqref{eq:Action} after integrating out the gauge fields is
\EQ{
&
S[g,\CF;\alpha]=-\frac{k}{\pi\lambda}\int d^2x\, \Tr\left\{(1-\lambda)\, j_+^\mathfrak{f} j_-^\mathfrak{f}+ \alpha(1-\lambda)\, j_+^\perp j_-^\perp  +\frac{\lambda}{2} \,\CF^{-1}\partial_+\CF\, \CF^{-1}\partial_-\CF \right.\\[5pt]
&\quad 
\left.+\left((1-\lambda)j_+^\mathfrak{f} - \lambda\CF^{-1}\partial_+\CF\right)\, \left(\lambda \AD_{\CF}-1\right)^{-1} 
\left((1-\lambda)j_-^\mathfrak{f} + \lambda\partial_-\CF \CF^{-1}\right)  \right\}+S_\text{WZ}[\CF],
\label{eq:S-lambda-sub}
}
where $S_\text{WZ}$ is the Wess--Zumino term at level $k$. 
Because of the procedure of integrating out the gauge fields, we may actually identify a ``dilaton term'' in the action, similarly to what happens in the $\lambda$-deformed model \cite{Sfetsos:2014cea}. From path-integral arguments, we then find that the ``dilaton'' is given by
\EQ{\label{ScalarField}e^{-2\Phi}=e^{-2\Phi_0} k^{\text{dim}{\cal F}}\det\left(\lambda^{-1}-\text{Ad}_{\cal F}\right)\,,}
and $\Phi_0$ is the ``dilaton'' of the original model {(which is just a constant in the case of the PCM considered here)}.
Notice that the action~\eqref{eq:S-lambda-sub} inherits the previous gauge invariance, so that we have the gauge transformations
\EQ{
g\to U g,\qquad
\CF\to U\CF U^{-1},
\qquad U\in F.
\label{eq:GaugeTranf-lambda-sub}
}
When $F=G$, as discussed in~\cite{Sfetsos:2013wia}, the gauge invariance is enough to gauge fix $g=1$, so that one obtains an action for $\CF$ only, and  that is the one of the standard $\lambda$-deformed model 
\EQ{S[\CF]=S_\text{WZW}[\CF]+\frac{\lambda k}{\pi}\int d^2x \Tr\left\{\CF^{-1}\partial_+\CF\left(\lambda \AD_{\cal F}-1\right)^{-1}\partial_-\CF \CF^{-1}\right\}.\label{action.lambda.Sf}
}
If $F$ is a strict subgroup of $G$, however, the gauge choice $g=1$ is not possible. What is possible is to first parameterise
\EQ{
g= f q,\qquad f\in F,\qquad q\in F\backslash G
\label{eq:Param}
}
so that under the gauge transformation only $f$ changes as $f\to Uf$, while $q$ remains invariant. The gauge transformation can then be used to set 
\EQ{
f=1 \qquad \Rightarrow \qquad g= q\in F\backslash G.
}
Notice that this gauge choice does not necessarily imply any condition for the projections of $j=dgg^{-1}=dqq^{-1}$ on $\mathfrak{f}$ or on $\mathfrak{f}^\perp$, meaning that in general both of them may be non-vanishing. Notice that because $q$ is an element of the coset $F\backslash G$, it is parameterised by dim$(G)-$dim$(F)$ degrees of freedom, and together with the dim$(F)$ degrees of freedom in $\CF$, we have a total of dim$(G)$ as already remarked. The coset degrees of freedom in $q$ are what we call the ``spectator fields''.

\section{Symmetries}\label{sec:sym}
We have already discussed the local symmetries of the action that, in particular, act on the field $g$ from the left. In general, we do not expect more symmetries acting from the left{, with the exception of course of a possible global symmetry corresponding to a subgroup of $G$ that commutes with $F$}. There is still, of course, a \emph{global} symmetry that transforms only the field $g$
\EQ{
g\to g\, M,\qquad M\in G,
\label{eq:Global}
}
by acting from the \emph{right} with an element of the \emph{full} group $G$. This symmetry should be treated with care after the gauge fixing $g=q$ discussed in the previous section. In fact, suppose that we fixed the gauge so that $q=\exp(X)$ with $X\in \mathfrak{f}^\perp$. Now a global right $G$ action may spoil the gauge choice, so that one needs a compensating \emph{local} left transformation to go back to the desired gauge slice. In other words, we need
\EQ{
q\to q'= U_{[M]}\ q\ M:= \exp(X'),\qquad M\in G,\qquad U_{[M]}\in F,
}
where $U_{[M]}$ must be chosen so that $X'\in \mathfrak{f}^\perp$. Because of this compensating gauge transformation, we conclude that after the gauge fixing  the global $G$ symmetry acts as
\EQ{
q\to q'= U_{[M]}\ q\ M,\qquad
\CF\to U_{[M]}\ \CF\ U_{[M]}^{-1}.
}
In general, $U_{[M]}$ may depend  on $M$ in a complicated way. One exception is when we restrict $M$ to take value just in $F$. In that case, we see that it is enough to take $U_{[M]}=M^{-1}$ because then
\EQ{
q'=\exp(X')=M^{-1}\exp(X)M=\exp(M^{-1}XM).
}
In fact, the transformation $X'=M^{-1}XM$ is compatible with the gauge choice $X'\in \mathfrak{f}^\perp$ thanks to~\eqref{eq:comm-f-fperp}. Notice that in this particular case the matrix $U_{[M]}$ implementing the gauge transformation turns out to be constant. We conclude that the subgroup $F$ has a simple vectorial action on both fields
\EQ{
q\to q'= M^{-1}\ q\ M,\qquad
\CF\to M^{-1}\ \CF\ M.
}
This is a global gauge transformation, and we expect that  these theories admit soliton solutions which carry kink charges associated to these symmetry transformations like in~\cite{Hollowood:2010dt,Hollowood:2011fq,Hollowood:2013oca}. Let us emphasise that $F$ is just the subgroup of the global symmetry that acts without isotropy, but as explained above there is in general a larger $G$ global symmetry that should give rise to regular Noether charges. The case of the standard $\lambda$-deformation with $G=F$ is simpler. Because of the allowed gauge choice $g=q=1$, it is obvious that we can take $U_{[M]}=M^{-1}$ and, thus, we recover the global symmetry discussed in~\cite{Sfetsos:2013wia} acting simply as $\CF\to M^{-1}\CF M$. 

\vspace{12pt}

Let us now discuss two discrete symmetries of the model. First, the action~\eqref{eq:S-lambda-sub} and the ``dilaton''~\eqref{ScalarField} are invariant under the generalised parity symmetry
\EQ{\sigma^+\leftrightarrow\sigma^-\,,\quad \CF\to\CF^{-1}\,,\quad 
(g,\lambda,\alpha)\to (g,\lambda,\alpha)\,.\label{Eq:Parity.Symmetry}}
The standard $\lambda$-deformation of~\cite{Sfetsos:2013wia} is also known to be invariant under a ``non-perturbative $\mathbb Z_2$ symmetry'' implemented as~\cite{Sfetsos:2014jfa,Georgiou:2015nka}
\EQ{
k\to -k,\quad
\lambda\to \lambda^{-1},\quad
\CF\to \CF^{-1}.
\label{eq:sym-alpha0}
}
We see that we can generalise this symmetry transformation also for the action~\eqref{eq:S-lambda-sub} if we additionally demand that under the $\mathbb{Z}_2$ transformation
\EQ{
q\to q,\qquad \alpha\to \lambda^{-1}\alpha.
\label{eq:sym-alpha}
}
It is perhaps easier to check when rewriting the action~\eqref{eq:S-lambda-sub} in the alternative form
\EQ{
&
S=S_\text{WZ}[\CF]-\frac{k}{\pi}\int d^2x\, \Tr\left\{\frac12\CF^{-1}\partial_+\CF\left(1+\lambda^{-1}\AD_\CF^{-1}\right)\left(\lambda^{-1}-\AD_\CF\right)^{-1}\partial_-\CF \CF^{-1} \right.\\[5pt]
&\quad 
+(\lambda^{-1}-1)\, j_+^\mathfrak{f}\left(\AD_\CF-1\right)\left(\AD_\CF-\lambda^{-1}\right)^{-1} j_-^\mathfrak{f} + \alpha(\lambda^{-1}-1)\, j_+^\perp j_-^\perp \\[5pt]
&\quad 
\left.+(\lambda^{-1}-1)j_+^\mathfrak{f}\left(\AD_\CF-\lambda^{-1}\right)^{-1}\partial_-\CF \CF^{-1} 
-(\lambda^{-1}-1)\CF^{-1}\partial_+\CF\left(\AD_\CF-\lambda^{-1}\right)^{-1}j_-^\mathfrak{f}
 \right\}.
 \label{eq:alt-S-lambda-sub}
}
Notice that the ``dilaton''~\eqref{ScalarField} is invariant under this  $\mathbb{Z}_2$ transformation if we also transform
\EQ{
\Phi_0\to \Phi_0+\frac12  \log\lambda\ \text{dim}\CF.
}
In section~\ref{sec:int}  we will show that, in most cases of interest integrability fixes the parameter $\alpha$ to be  
\EQ{
\alpha=\frac{1+\lambda}{2},
\label{eq:alpha}
}
which indeed satisfies the above property in~\eqref{eq:sym-alpha}. Therefore, we conclude that the $\lambda$-deformed action is invariant under the $\mathbb{Z}_2$ symmetry
\EQ{
k\to -k,\qquad
\lambda\to \lambda^{-1},\qquad
\CF\to \CF^{-1},\qquad
q\to q.
}
The only cases in which $\alpha$ can be generic are the case of $F$ abelian and the trivial case when $G=F\times F^\perp$ is of direct product form (i.e.~$[\mathfrak{f},\mathfrak{f}^\perp]=0$).

\section{Undeformed limits}\label{sec:lim}
Let us now discuss some special limits of the action~\eqref{eq:S-lambda-sub}. We will discuss in particular the limits $\lambda\to 0$ and $\lambda\to 1$, and we will start by showing that taking the naive limits leads to inconsistent results. For example, let us consider the limit $\lambda\to 0$. Looking at~\eqref{eq:Couplings} we see that if we wanted to keep $\kappa^2$ finite, then we should also take $k\to 0$, which means that we would explicitly loose the $\text{dim}(F)$ degrees of freedom coming from the gWZW part of the action.
In this naive limit, in fact, we would obtain just
\EQ{
S\to -\frac{\kappa^2}{\pi}\int d^2x\, \Tr\left\{\alpha^*\, j_+^\perp j_-^\perp\right\},
\label{eq:naive-lambda0}
}
where $\displaystyle\alpha^*=\lim_{\lambda\to 0}\alpha$. Notice that when $\alpha$ is given by~\eqref{eq:alpha} then $\alpha^*=\nicefrac12$. The above is the action of the $F\backslash G$ coset space $\sigma$-model, so that only dim$(G)-$dim$(F)$ degrees of freedom survive.
Keeping, instead, $k$ fixed  implies that $\kappa^2\to \infty$ and the action appears not be finite. This conclusion, however, relies on the assumption that field configurations are independent of~$\lambda$. In contrast, following~\cite{Sfetsos:1994vz} and~\cite{Sfetsos:2013wia}, 
the proper definition of the limit $\lambda\to 0$ (and $\lambda\to 1$ as explained later) can be obtained by restricting to the field configurations that contribute to the path integral corresponding to our action, which  
exhibit a specific dependence on $\lambda$.  
Let us consider the path integral in the limit $\lambda\to0$ when $k$ is fixed and $\kappa^2\approx\nicefrac{k}{\lambda}\to \infty$. Then, the non-trivial contributions come from configurations close to an arbitrary constant element $g=g_0$ so that $\Tr\{j_+^\perp j_-^\perp\}\sim \mathcal{O}(\lambda)$, as required by demanding that the action~\eqref{eq:S-lambda-sub} is finite. For field configurations so that $\Tr\{j_+^\perp j_-^\perp\}\sim$ finite, the action becomes very large and, thus, their contribution is negligible. In our case, this leads us to consider only configurations {that, up to gauge transformations, are} of the form
\EQ{
g\approx \left(1+\sqrt{\lambda}\, \phi +\ldots\right)g_0, \quad \phi\in\mathfrak{g}\qquad \Rightarrow\qquad j_\pm = \partial_\pm g g^{-1}\approx \sqrt{\lambda}\, \partial_\pm\phi.
\label{eq:lambdaZeroLimit}
}
Note that, under the global transformation~\eqref{eq:Global}, we can transform $g_0\to g_0 M$, and therefore set $g_0=1$. In summary, we learn that we must take  the correlated limit
\EQ{
\lambda\to0, \qquad j_\pm\to \sqrt{\lambda}\; \partial_\pm \phi,
\label{limit.lambda.zero}
}
so that
\EQ{
S\to S_\text{WZW}[\CF] -\frac{k}{\pi}\int d^2x\, \Tr\left\{\alpha^*\, \partial_+ \phi
^\perp \,\partial_- \phi^\perp\right\}.
\label{eq:lambda-to-zero-1}
}
Here, $S_\text{WZW}$ is the WZW action for the group $F$ with (finite) level~$k$, and the second term is the action of 
a decoupled free boson taking values in $\mathfrak{f}^\perp$ with positive kinetic term.\footnote{The generators of the Lie algebra $\frak{g}$ are taken to be anti-Hermitian.\label{Conventions.algebra}} We see that in this limit the interactions in the coset $F\backslash G$ part of the action are suppressed, therefore the fact that $\phi^\perp$ takes values in $\mathfrak{f}^\perp$ is irrelevant, and the only important point is that there are $\text{dim}(G)-\text{dim}(F)$ free bosons. Altogether, the number of degrees of freedom is $\text{dim}(G)$ as desired.

By keeping the first orders in $\lambda$ one can understand the leading  term \eqref{eq:lambda-to-zero-1} of the action~\eqref{eq:S-lambda-sub} obtained in this $\lambda\to 0$ limit. One finds
\EQ{
S\to &S_\text{WZW}[\CF] -\frac{k}{\pi}\int d^2x\, \Tr\left\{\alpha^*\, \partial_+ \phi
^\perp \,\partial_- \phi^\perp\right\}\\
&-\frac{k}{\pi}\int d^2x\, \Tr\left\{\sqrt{\lambda}\left(\CF^{-1}\partial_+\CF \partial_-\phi^{\mathfrak{f}}-\partial_+\phi^{\mathfrak{f}}\partial_-\CF\CF^{-1}\right)\right.\\
&\qquad\qquad \left.+\lambda \partial_+\phi^{\mathfrak{f}}\left(1-\AD_{\CF}\right)\partial_-\phi^{\mathfrak{f}}
+\lambda \CF^{-1}\partial_+\CF\partial_-\CF\CF^{-1}\right\}+\mathcal O(\lambda^{\nicefrac32}).
}
Notice that the requirement of gauge invariance  mixes terms at different orders of $\lambda$. In the case of the standard $\lambda$ deformation (when $G=F$) we can fix the gauge so that $\phi^\perp=\phi^\mathfrak{f}=0$, and we recover the result of~\cite{Sfetsos:2013wia}; namely  that the perturbation is given by $ -\lambda\,\frac{k}{\pi}\int d^2x\, \Tr\left\{\mathcal J_+\bar{\mathcal  J}_-\right\}$, with $\mathcal J_+=\CF^{-1}\partial_+\CF$ and $\bar{\mathcal  J}_-=\partial_-\CF\CF^{-1}$.

Obviously, even when discussing the $\lambda\to 0$ limit, one should take into account the gauge invariance of the action. The above discussion, then, can be slightly generalised, meaning that instead of~\eqref{eq:lambdaZeroLimit} we can take 
\EQ{
g\approx U\left(1+\sqrt{\lambda}\, \phi +\ldots\right)g_0, 
\label{eq:lambdaZeroLimit-2}
}
where $U\in F$ can be field-dependent, and where we do not need to specify the $\lambda$-dependence of $U$. In particular, it may be non-trivial in the $\lambda\to 0$ limit. In fact, even if $U\neq 1$, we can always implement a gauge transformation $g\to U^{-1}g,\CF\to U^{-1}\CF U$ to reabsorb the $U$-dependence. When $U\neq 1$, it is easy to check that, comparing to the previous formulas, we now have $j^\perp_\pm\to Uj^\perp_\pm U^{-1}$ and $j^{\mathfrak{f}}_\pm\to Uj^{\mathfrak{f}}_\pm U^{-1}+\partial_\pm UU^{-1}$, so that now  $j^{\mathfrak{f}}_\pm$ is not necessarily $\mathcal{O}(\sqrt{\lambda})$, and it can be $\mathcal{O}(1)$.\footnote{Notice that this does not spoil the well-behaviour of the action in the $\lambda\to 0$ limit, because the leading order of the $j^{\mathfrak{f}}_+j^{\mathfrak{f}}_-$ term in the action vanishes, thanks to a cancellation between the first and the second line in~\eqref{eq:S-lambda-sub}.} Therefore, in this more general setup, the leading behaviour of the action in the $\lambda\to 0$ limit reads
\EQ{
&
S\to S_\text{WZW}[\CF] -\frac{k}{\pi}\int d^2x\, \Tr\left\{\alpha^*\, \partial_+ \phi
^\perp \,\partial_- \phi^\perp+\partial_+UU^{-1}(1-\AD_\CF)\partial_-UU^{-1}\right.\\
&\qquad\qquad\qquad\qquad\left.+\CF^{-1}\partial_+\CF \partial_-UU^{-1}-\partial_+UU^{-1}\partial_-\CF\CF^{-1}\right\}.
\label{eq:lambda-to-zero-2}
}
Even though the action~\eqref{eq:lambda-to-zero-2} looks different from~\eqref{eq:lambda-to-zero-1}, it is important to stress that it is equivalent thanks to gauge transformations. In particular, the interpretation of the model given above in the $\lambda\to 0$ limit (namely a WZW on $F$ plus $\text{dim}(G)-\text{dim}(F)$ free bosons) is still valid.

The discussion for the $\lambda\to 1$ limit is similar. From~\eqref{eq:Couplings}, if we keep $k$ fixed then $\kappa^2\to0$, which means that we loose the $\text{dim}(G)-\text{dim}(F)$ degrees of freedom coming from the gPCM part of the action. In this naive limit, in fact, we would get 
\EQ{
S\to S_\text{WZW}[\CF]+\frac{k}{\pi}\int d^2x\, \Tr\left\{\CF^{-1}\partial_+\CF\, \left(\AD_\CF-1\right)^{-1} 
\partial_-\CF \CF^{-1} \right\},
}
where $S_\text{WZW}$ is the Wess--Zumino--Witten action at level~$k$. {This action is ill-defined as it is divergent, as $\text{Ad}_\CF$ is an orthogonal matrix and one of its eigenvalues is 1. In addition,}  it incorporates only dim$(F)$ degrees of freedom. Here we assumed that $\alpha$ remains finite when $\lambda\to 1$ (in fact, it goes to 1 for~\eqref{eq:alpha}). 
Instead, we will keep $\kappa^2$ fixed, which implies that $k\to\infty$, and to avoid that the WZW part of the action becomes infinite we have to restrict the field configurations appropriately. In particular, as in~\cite{Sfetsos:2013wia}, we take the correlated limit
\EQ{
\lambda\to 1, \qquad \CF\to 1+ (1-\lambda)\, \nu+\ldots, \qquad k(1-\lambda)=\kappa^2=\text{ finite},
\label{limit.lambda.one}
}
so that the action~\eqref{eq:Action} becomes
\EQ{
S\to -\frac{\kappa^2}{\pi}\int d^2 x\, \Tr\left\{\left(j_+ + A_+\right)^{\mathfrak{f}}\, \left(j_- + A_- \right)^{\mathfrak{f}}+\tilde\alpha\, j_+^\perp j_-^\perp+ \nu F_{+-}\right\},
\label{eq:NATDaction1}
}
up to a boundary term and where we have defined $\displaystyle\tilde\alpha=\lim_{\lambda\to 1}\alpha$.
Since $\nu\in \mathfrak{f}=\mathfrak{f}^\ast$, this is the starting point to construct the non-Abelian T-duality action with respect to the subgroup $F\in G$ (see for example~\cite{Sfetsos:2013wia,Borsato:2016pas}).  Notice, in fact, that in the $\lambda\to 1$ limit the solutions~\eqref{eq:EqsGaugeFields} agree with~\cite{Borsato:2016pas}, namely
\EQ{
A_\pm=\pm (1\pm \Ad_\nu)^{-1}\left(\partial_\pm \nu\mp j_\pm^{\mathfrak{f}}\right).
}
Inserting the latter into \eqref{eq:NATDaction1} we find the T-dual action
\EQ{S=-\frac{\kappa^2}{\pi}\int d^2x\Tr\left\{(\partial_+v -j_+^\mathfrak{f})(1-\Ad_v)^{-1}(\partial_-v+j_-^\mathfrak{f})+j^\mathfrak{f}_+j^\mathfrak{f}_-+\tilde\alpha j_+^\perp j_-^\perp\right\}\,,}
that is invariant under the generalised parity symmetry
$\sigma\to-\sigma$ and $v\to-v$. The latter action can be directly obtained from \eqref{eq:S-lambda-sub} using \eqref{limit.lambda.one}.

To conclude, in the two limits we find
\EQ{
\lambda\to 0 \quad(\kappa^2\to \infty): & \qquad \text{WZW on }F \text{ and dim}(G)-\text{dim}(F) \text{ free bosons}, \\
\lambda\to 1 \quad(k\to \infty)\;:& \qquad \text{NATD of PCM on }G \text{ where only }F \text{ is dualised}.}
Note that in each limit we only have to restrict the set of allowed configurations for one of the fields.
Interestingly, in both limits we recover integrable models, and this is true without imposing additional constraints, for example, on the algebraic structure of $F$ and $G$. A natural hope is to construct an integrable deformation interpolating between these two limits. In the next section we will show that this is not always possible, and that we can write a Lax connection only when requiring an extra algebraic structure.

For the $\lambda$-deformed model \eqref{action.lambda.Sf} in the parametric space $(\lambda,k)$ there exists another interesting correlated limit as $\lambda\to-1$ and $k\gg1$~\cite{Georgiou:2016iom}.\footnote{Note that the value $\lambda=-1$ cannot be realised from the definition of $\lambda$ in \eqref{eq:Couplings} for the signs of the couplings $\kappa$ and $k$ but it makes sense at the level of the action \eqref{action.lambda.Sf}.}  In particular, the limit requires a zoom-in procedure analogue to that of the non-abelian T-dual leading to the pseudo-dual model~\cite{Nappi:1979ig}. The generalisation of this limit for~\eqref{eq:alt-S-lambda-sub} can be straightforwardly worked out along the lines of the $\lambda\to0$ limit where in this case
\EQ{\lambda=-1+\frac{\xi}{k^{\nicefrac{1}{3}}}\,,\quad \alpha\to\frac{\alpha}{k^{\nicefrac{1}{3}}}\,,\quad
\CF=1+\frac{v}{k^{\nicefrac{1}{3}}}\,,\quad
g=U\left(1+\frac{\phi}{k^{\nicefrac{1}{3}}}\right)g_0\,,\quad k\gg1\,,
}
with $U\in\CF$, $v\in\frak{f}$ and $\phi\in\frak{g}$.
Inserting the above into \eqref{eq:alt-S-lambda-sub} we find the action
\EQ{S=&
\frac1\pi\int d^2x\,\Tr\left\{
-\frac{1}{4}\partial_+v\left(\xi+\frac13\Ad_v\right)\partial_-v+2\alpha\partial_+\phi^\perp\partial_-\phi^\perp\right.\\
&\left.+\partial_+U U^{-1}\Ad_v\partial_-UU^{-1}
-k^{\nicefrac13}\left(\partial_+v\partial_-U U^{-1}-\partial_+UU^{-1}\partial_-v\right)
\right\}\,.}
The interpretation of the above action up to gauge transformations is the pseudo-dual model plus $\dim(G)-\dim(F)$ free boson of opposite kinetic term (see footnote~\ref{Conventions.algebra}).

\section{Comments on the degenerate case}\label{sec:deg}
As emphasised in section~\ref{sec:action}, in our construction we assume that the restriction to $\mathfrak{f}$ of the bilinear form on $\mathfrak{g}$ is non-degenerate. 
This assumption is, in fact, more restrictive than those needed to apply NATD which, as explained above, is recovered in the limit $\lambda\to1$.
We can work out the conditions that enable the application of NATD to the PCM with a subgroup $F\subset G$ by looking at the action~\eqref{eq:NATDaction1}
\EQ{
S\to -\frac{\kappa^2}{\pi}\int d^2 x\, \left[\langle j_+ + A_+\,,\,  j_-+ A_- \rangle 
+ \langle \nu, F_{+-}\rangle\right].
}
We have written the action in terms of a general non-degenerate,
symmetric, and ad-invariant bilinear form $\langle \,, \rangle$ that is not necessarily the trace form associated to a matrix representation of $\mathfrak{g}$.
Here, $\nu \in\mathfrak{f}^\ast$ and, in general, $\mathfrak{f}^\ast\not= \mathfrak{f}$. Note that the definition of the dual $\mathfrak{f}^\ast$ depends on the choice of the bilinear form. The NATD action is obtained by integrating out the gauge fields $A_\pm$ which, classically, is equivalent to imposing their equations of motion
\EQ{
\text{P}^{\mathfrak{f}^\ast}\left( \left(1 \pm \text{ad}_\nu\right)A_\pm\right)=\pm\partial_\pm \nu -j_\pm^\ast,
\label{eq:NATDequation}
}
where $\text{P}^{\mathfrak{f}}$ and $\text{P}^{\mathfrak{f}^\ast}$ are projectors on $\mathfrak{f}$ and $\mathfrak{f}^\ast$, respectively, and $\text{ad}_\nu(\cdot)=\left[\nu, \cdot\right]$. For the construction to work, one should be able to use these equations  to write $A_\pm$ in terms of~$\partial_\pm\nu$ and $j_\pm^\ast$. This translates into the condition that the {linear} operator ${\cal O}= \text{P}^{\mathfrak{f}^\ast}\left(1-\text{ad}_\nu \right) \text{P}^{\mathfrak{f}}$ must be invertible, which requires in particular that each element in $\mathfrak{f^\ast}$ is the image of an element in $\mathfrak{f}$. Since $\mathfrak{f}$ and $\mathfrak{f}^\ast$ are of the same dimension, this is equivalent  to requiring that ker$(\mathcal{O})=\{0\}$. 
When the restriction of the non-degenerate bilinear form of $G$ to  $F$ is non-degenerate, like in our case, then $\mathfrak{f}^\ast= \mathfrak{f}$ and the condition that ker$(\mathcal{O})=\{0\}$ is obviously satisfied in a neighbourhood of $\nu=0$. In contrast, when the restriction of the non-degenerate bilinear form of $G$ to  $F$ is degenerate this condition becomes non trivial and, in particular, it is not satisfied at $\nu=0$. If $\mathcal{O}$ is invertible, it means that the invertibility is consequence of the $\Ad_\nu$ in its definition. Therefore, even if the NATD procedure can be applied, the resulting action will be singular at $\nu=0$. 

The fact  that the restriction of the non-degenerate bilinear form of $G$ to  $F$ may be degenerate is not easily translatable into a condition on the Lie algebraic structure of~$F$.
On the one hand, when $\mathfrak{g}$ and, thus, $\mathfrak{f}$ are realised as a faithful representation in terms of matrices and the bilinear form is the one induced by the trace of their products, Cartan's criteria ensure that $\mathfrak{f}$ is non-semisimple if the bilinear form of $\mathfrak{f}$ is degenerate (see, for example,~\cite{Jacobson:1979ja} sec. III-4).\footnote{In turn, it cannot be ensured that $\mathfrak{f}$ is semisimple when the 
bilinear form of $G$ to $F$ is non-degenerate, unless of course the bilinear form is the Killing one.}
However, this is far form being the only possibility. An interesting example is provided by the so-called complex Drinfel'd double associated to a compact real simple algebra $\mathfrak{f}$ (see~\cite{Vicedo:2015pna,Hoare:2017ukq} for rather detailed descriptions). Let $\mathfrak{f}^{\mathbb C}$ be the complexification of $\mathfrak{f}$ and introduce a Cartan--Weyl basis composed of the Cartan generators $\{h_i\}$ and positive $\{e_M\}$ and negative
$\{f_M\}$ roots. The Cartan generators and the simple roots satisfy the defining relations
\EQ{
\left[h_i,e_j\right]= a_{ij} e_j,\qquad 
\left[h_i,f_j\right]= -a_{ij} f_j, \qquad
\left[e_i,f_j\right]= \delta_{ij}\, h_j,
}
where $a_{ij}$ is the symmetrised Cartan matrix, and the Killing form $\kappa(x\, y)= \Tr\left(\text{ad}_x\, \text{ad}_y\right)$ can be normalised such that
\EQ{
\kappa\left(h_i\, h_j\right)= a_{ij}, \qquad
\kappa\left(e_M\, f_N\right)= \delta_{MN}.
}
The algebra $\mathfrak{g}=\mathfrak{f}^{\mathbb C}$ can be regarded as a real Lie algebra, and it can be decomposed into two real Lie subalgebras $\mathfrak{g}= \mathfrak{f}+\tilde{\mathfrak{f}}$.
Here, $\mathfrak{f}$ is the compact real form itself generated by
\EQ{
\{t_a\}= \left\{i h_j, i(e_M + f_M), (e_M-f_M) \right\},
}
and $\tilde{\mathfrak{f}}$ is the Borel subalgebra generated by
\EQ{
\{t^a\}= \left\{h_j, e_M, i e_M \right\}.
}
The real algebra $\mathfrak{g}$ admits the symmetric, ad-invariant, and non-degenerate bilinear form
\EQ{
\langle x,y\rangle = \text{Im}\,\kappa(x,y), \qquad x,y\in \mathfrak{g},
}
which satisfies
\EQ{
\langle t_a,t^b\rangle= \delta_{a}{}^b,\qquad \langle t_a,t_b\rangle=\langle t^a,t^b\rangle=0.
}
Note that $\langle\,,\rangle$ is not the trace form of any representation of~$\mathfrak{g}$. Therefore, in this case the restriction of the non-degenerate bilinear form $\langle\,,\rangle$ to $\mathfrak{f}$ is completely degenerate (both $\mathfrak{f}$ and $\tilde{\mathfrak{f}}$ are maximally isotropic), but the subalgebra $\mathfrak{f}$ is simple. Moreover, $\tilde{\mathfrak{f}}$ is the dual of $\mathfrak{f}$ with respect to this bilinear form, and it is a solvable Lie algebra.\footnote{Using that $\mathfrak{f}$ is simple in this particular example, it can be checked that the operator ${\cal O}$ defined after~\eqref{eq:NATDequation} is actually invertible at $\nu\not=0$.}

The major obstacle to generalise our construction  is that one can define the WZW action in terms of $F$ only if it is equipped with a bilinear form that is symmetric, ad-invariant, and non-degenerate. {This obstruction makes it difficult to generalise the construction to the case when the restriction on $\mathfrak{f}$ of the bilinear form on $\mathfrak{g}$ is degenerate.}
In the particular case of non-semisimple Lie groups, the crucial observation made in~\cite{Nappi:1993ie} is that some of them admit several bilinear forms that are symmetric and ad-invariant, and some of those bilinear forms may even be non-degenerate. Therefore, they can be used to construct a WZW action and, possibly, new generalisations of the $\lambda$-deformation. In~\cite{Sfetsos:2022irp}, this was used to construct a generalisation of the usual $\lambda$-deformation (with $F=G$).

\section{Equations of motion and Lax connection}\label{sec:int}

The equations of motion of the action~\eqref{eq:Action}
are derived in appendix~\ref{app:equations-of-motion}. In terms of the current
\EQ{
j_\pm = \partial_\pm g g^{-1} = j^\mathfrak{f}_\pm + j^\perp_\pm
}
and its components with respect to the decomposition $\mathfrak{g}= \mathfrak{f}\oplus \mathfrak{f}^\perp$, they can be written as
\AL{
&
\left[\partial_\pm + \frac{1-\lambda}{\lambda}\, j^\mathfrak{f}_\pm + \frac{1}{\lambda}\, A_\pm,\partial_\mp +A_\mp\right]=0,
\label{eq:eomF}\\[5pt]
&
\left[\partial_+ +\frac{1-\alpha}{\alpha}\, j^\mathfrak{f}_+ + \frac{1}{\alpha}\, A_+, j^\perp_-\right]+ 
\left[\partial_- +\frac{1-\alpha}{\alpha}\, j^\mathfrak{f}_- + \frac{1}{\alpha}\, A_-, j^\perp_+\right]=0,
\label{eq:eomf}
}
together with the Maurer--Cartan identity
\EQ{
\left[\partial_+ - j_+, \partial_- - j_-\right]=0,
\label{eq:MCidentity}
}
and the equations of motion of the gauge fields $A_\pm$
\EQ{
&
\frac{1-\lambda}{\lambda} \, \left(j^\mathfrak{f}_- +A_-\right)=\left(-\partial_-\CF\CF^{-1} +\CF A_-\CF^{-1}-A_-\right),\\[5pt]
&
\frac{1-\lambda}{\lambda}\, \left(j^\mathfrak{f}_+ +A_+\right)=\left(\CF^{-1}\partial_+\CF +\CF^{-1} A_+\CF-A_+\right).
\label{eq:eomANew}
}
The last two equations allow one to write the fields $A_\pm$ in terms of $j_\pm^\mathfrak{f}$ and $\CF$, and they will not be used in the rest of this section.

The equations~\eqref{eq:eomF} allow one to write $\partial_\pm A_\mp$ in terms of other field combinations as follows
\AL{
&
{\cal E}1:\quad \partial_+ A_-=-\frac{1}{1+\lambda}\, \bigg(\left[\partial_+ + A_+,j^\mathfrak{f}_-\right] + \left[\partial_- + A_-,j^\mathfrak{f}_+\right]\nonumber \\[5pt]
&
\qquad\qquad\qquad
+
\left[(1-\lambda)\, j^\mathfrak{f}_+ + A_+, A_-\right] - (1-\lambda)\, \partial_- j^\mathfrak{f}_+\bigg),\label{eq:EQder1}\\[5pt]
&
{\cal E}2:\quad \partial_- A_+=-\frac{1}{1+\lambda}\, \bigg(\lambda \left[\partial_+ + A_+,j^\mathfrak{f}_-\right] + \lambda \left[\partial_- + A_-,j^\mathfrak{f}_+\right]\nonumber \\[5pt]
&
\qquad\qquad\qquad
-
\left[(1-\lambda)\, j^\mathfrak{f}_+ + A_+, A_-\right] + (1-\lambda)\, \partial_- j^\mathfrak{f}_+\bigg).\label{eq:EQder2}
}
Similarly, eq.~\eqref{eq:eomf} and the components of the Maurer--Cartan identity on $\mathfrak{f}^\perp$ allow one to write
\AL{
&
{\cal E}3:\quad 2\partial_+ j_-^\perp=-\frac{1}{\alpha}\left[A_++(1-2\alpha)\, j_+^\mathfrak{f}, j_-^\perp\right]-\frac{1}{\alpha}\left[A_-+j_-^\mathfrak{f}, j_+^\perp\right]+\left[ j_+^\perp, j_-^\perp\right]^\perp,\label{eq:EQder3}
\\[5pt]
&
{\cal E}4:\quad 2\partial_- j_+^\perp=-\frac{1}{\alpha}\left[A_++j_+^\mathfrak{f}, j_-^\perp\right]-\frac{1}{\alpha}\left[A_-+(1-2\alpha)j_-^\mathfrak{f}, j_+^\perp\right]-\left[ j_+^\perp, j_-^\perp\right]^\perp.\label{eq:EQder4}
}
Finally, the components of the Maurer--Cartan identity on $\mathfrak{f}$ read
\EQ{
{\cal E}5:\quad \left[\partial_+ - j^\mathfrak{f}_+, \partial_- - j^\mathfrak{f}_-\right] +\left[ j_+^\perp, j_-^\perp\right]^\mathfrak{f}=0. \label{eq:EQder5}
}
These five equations are equivalent to the equations of motion~\eqref{eq:eomF} and~\eqref{eq:eomf} together with the Maurer--Cartan identity~\eqref{eq:MCidentity}.

The natural ansatz for the Lax connection is
\EQ{
\AA_\pm (z)= \Phi_\pm(z)\, A_\pm + \Psi_\pm(z)\,  j^\mathfrak{f}_\pm + \Upsilon_\pm(z)\,  j^\perp_\pm,
\label{eq:LaxAnsatz}
}
and it will be convenient to distinguish the components of $\left[\partial_+ +\AA_+(z), \partial_- +\AA_-(z)\right]$ on~$\mathfrak{f}$ and~$\mathfrak{f}^\perp$. They are given by the equations:
\AL{
\left[\partial_+ +\AA_+(z), \partial_- +\AA_-(z)\right]^\mathfrak{f}&=\Phi_-\partial_+ A_- +\Psi_- \partial_+ j^\mathfrak{f}_- -\Phi_+\partial_- A_+ -\Psi_+ \partial_- j^\mathfrak{f}_+\nonumber\\[5pt]
&\hspace{-1cm}
+\left[\Phi_+ A_+ + \Psi_+ j^\mathfrak{f}_+, \Phi_- A_- + \Psi_- j^\mathfrak{f}_-\right]
+\Upsilon_+\Upsilon_- \left[ j_+^\perp, j_-^\perp\right]^\mathfrak{f}
\label{eq:OnH}
}
and
\AL{
\left[\partial_+ +\AA_+(z), \partial_- +\AA_-(z)\right]^\perp&=\Upsilon_-\partial_+ j_-^\perp - \Upsilon_+\partial_- j_+^\perp
+\Upsilon_-\left[\Phi_+ A_+ + \Psi_+ j^\mathfrak{f}_+,  j_-^\perp\right]\nonumber\\[5pt]
&\qquad
+\Upsilon_+\left[ j_+^\perp,\Phi_- A_- + \Psi_- j^\mathfrak{f}_-\right] +\Upsilon_+\Upsilon_- \left[ j_+^\perp, j_-^\perp\right]^\perp.
\label{eq:OnHperp}
}
Using the equations~\eqref{eq:EQder1} and~\eqref{eq:EQder2} to eliminate $\partial_+ A_-$ and $\partial_-A_+$ in~\eqref{eq:OnH}, and imposing that all the remaining terms vanish, give rise to the following identities:
\EQ{
&
[A_+,j_-^\mathfrak{f}]\;\; \longrightarrow\;\; \Phi_- -\lambda\Phi_+- (1+\lambda)\Phi_+\Psi_-=0,\\[5pt]
&
[A_-,j_+^\mathfrak{f}]\;\; \longrightarrow\;\; 
\Phi_+-\lambda \Phi_- - (1+\lambda)\Phi_-\Psi_+=0,\\[5pt]
&
[A_+,A_-]\;\; \longrightarrow\;\; \Phi_++ \Phi_- - (1+\lambda)\Phi_+ \Phi_-=0,
\label{eq:problemBo}
}
where on the left-hand-side we have indicated the specific combination of fields that multiplies the coefficient that has to vanish. 
{The above equations should be imposed in the general case of $F$ non-abelian. In fact, when $F$ is abelian the commutators that they multiply are automatically zero, and we don't find constraints for the functions from these terms.} Assuming that $\Phi_+,\Phi_-\not=0$, these identities are equivalent to
\EQ{
\Phi_++ \Phi_- = (1+\lambda)\Phi_+ \Phi_-,\qquad
\Psi_\pm = \Phi_\pm -1.
\label{eq:solH}
}
The rest of the terms coming from~\eqref{eq:OnH} give rise to
\EQ{
&\Psi_+\Psi_-\left(\partial_- j_+^\mathfrak{f}
-\partial_+ j_-^\mathfrak{f}
+\left[ j_+^\mathfrak{f},  j_-^\mathfrak{f}\right]\right)
+\Upsilon_+\Upsilon_- \left[ j_+^\perp,  j_-^\perp\right]^\mathfrak{f}=\\[5pt]
&\hspace{4cm}
=\left(\Upsilon_+\Upsilon_- - \Psi_+\Psi_-\right)\left[ j_+^\perp,  j_-^\perp\right]^\mathfrak{f}=0,
\label{eq:mcH}
}
where we have used~\eqref{eq:EQder5}.

Similarly, using the equations~\eqref{eq:EQder3} and~\eqref{eq:EQder4} to eliminate $\partial_+ j^\perp_-$ and $\partial_-j^\perp_+$ in~\eqref{eq:OnHperp}, and imposing that all the remaining terms vanish give rise to
\EQ{
&
[A_+,j_-^\perp]\;\; \longrightarrow\;\; 
\Upsilon_+-\Upsilon_- +2\alpha\,\Phi_+\Upsilon_-=0,\\[5pt]
&
[A_-,j_+^\perp]\;\; \longrightarrow\;\; \Upsilon_+-\Upsilon_- -2\alpha\, \Phi_-\Upsilon_+=0,\\[5pt]
&
[j_+^\mathfrak{f},j_-^\perp]\;\; \longrightarrow\;\; \Upsilon_+ +(2\alpha-1)\,\Upsilon_- + 2\alpha\, \Psi_+\Upsilon_-=0,\\[5pt]
&
[j_-^\mathfrak{f},j_+^\perp]\;\; \longrightarrow\;\; (2\alpha-1)\, \Upsilon_+ +\Upsilon_- + 2\alpha\,  \Psi_-\Upsilon_+=0,\label{eq:solHperp}
}
{which are equivalent to
\EQ{
\Phi_+ +\Phi_-= 2\alpha\, \Phi_+\Phi_-,\qquad
\Psi_\pm= \Phi_\pm -1, \qquad \Upsilon_-=-\frac{\Phi_-}{\Phi_+}\, \Upsilon_+\,,
\label{eq:solHperpUpsilon}
}
if we assume that $\Upsilon_+,\Upsilon_-\not=0$, and we also have}
\EQ{
\left(\frac{1}2 \Upsilon_+ +\frac{1}2 \Upsilon_- + \Upsilon_+\Upsilon_-\right) \, \left[ j_+^\perp, j_-^\perp\right]^\perp=0.
\label{eq:mcHperp}
}
{
To summarise, we have the equations collected in the Table~\ref{tab:int}.

\begin{table}
    \centering
    \begin{tabular}{|c|c|}
    \hline
        $[\mathfrak{f},\mathfrak{f}]$ & $\Phi_++ \Phi_- = (1+\lambda)\Phi_+ \Phi_-,\qquad
\Psi_\pm = \Phi_\pm -1$\\
         \hline
$[\mathfrak{f}^\perp,\mathfrak{f}^\perp]^{\mathfrak{f}}$ 
         & $\Upsilon_+\Upsilon_- =\Psi_+\Psi_-$\\
         \hline
$[\mathfrak{f},\mathfrak{f}^\perp]$ & $\Phi_+ +\Phi_-= 2\alpha\, \Phi_+\Phi_-,\qquad
\Psi_\pm= \Phi_\pm -1, \qquad \Upsilon_-=-\frac{\Phi_-}{\Phi_+}\, \Upsilon_+$\\
\hline
$[\mathfrak{f}^\perp,\mathfrak{f}^\perp]^{\perp}$ & $\frac{1}2 \Upsilon_+ +\frac{1}2 \Upsilon_- + \Upsilon_+\Upsilon_-=0$\\
\hline
    \end{tabular}
    \caption{Equations for integrability.}
    \label{tab:int}
\end{table}

In the generic case, $F$ is non-abelian and $\mathfrak{f}$ does not commute with $\mathfrak{f}^\perp$  so that we have to take into account the equations in the first and third lines of the table.
They are compatible only for the specific value 
\EQ{
\alpha=\frac{1+\lambda}{2}.
\label{eq:FixingAlpha}
}
}
{Then, the result is that the action~\eqref{eq:Action} can only be classically integrable for this particular value of $\alpha$
with Lax connection}
\EQ{
\AA_\pm (z)= \Phi_\pm(z)\, \left(A_\pm + j^\mathfrak{f}_\pm \right) -j^\mathfrak{f}_\pm+ \Upsilon_\pm(z)\, j^\perp_\pm,
}
where $\Phi_\pm$ satisfy
\EQ{
\Phi_+ +\Phi_- =(1+\lambda)\Phi_+\Phi_-
\label{eq:TheEquation}
}
and $\Upsilon_\pm$ satisfy eqs.~\eqref{eq:mcH} and~\eqref{eq:mcHperp} {(second and fourth lines of the table)} with
\EQ{
\Psi_\pm= \Phi_\pm -1, \qquad \Upsilon_-=-\frac{\Phi_-}{\Phi_+}\, \Upsilon_+.
\label{eq:solHperpUpsilonB}
} 
Always assuming that $\mathfrak f$ is non-abelian and that it does not commute with $\mathfrak f^\perp$, the solutions to the equations~\eqref{eq:mcH} and~\eqref{eq:mcHperp} depend on the algebraic structure of $[\mathfrak{f}^\perp,\mathfrak{f}^\perp]$, and one can distinguish three different cases
(the second case turns out to be an exception where $\mathfrak f$ actually commutes with $\mathfrak f^\perp$).

\subsection{No algebraic condition: $\lambda=1$ and NATD}
The first possibility would be to solve the two equations~\eqref{eq:mcH} and~\eqref{eq:mcHperp} without making any assumption about $[\mathfrak{f}^\perp, \mathfrak{f}^\perp]$. In other words, we simply impose
\EQ{
&
\Upsilon_+\Upsilon_- - \Psi_+\Psi_-=0,\\[5pt]
&
\frac{1}2 \Upsilon_+ +\frac{1}2 \Upsilon_- + \Upsilon_+\Upsilon_-=0,
\label{eq:General}
}
with
\EQ{
\Psi_\pm=\Phi_\pm-1,\qquad 
\Upsilon_-=-\frac{\Phi_-}{\Phi_+}\, \Upsilon_+,\qquad \Phi_+ +\Phi_- =(\lambda+1)\Phi_+\Phi_-.
}
It is easy to check that these equations are compatible only for $\lambda=1$.

To make the form of the solution more concrete, we will consider the usual resolution of the last equation in the context of the PCM and its $\lambda$-deformation:
\EQ{
\Phi_\pm(z)=\frac{z}{z\pm 1}\, \frac{2}{1+\lambda}.
\label{eq:standard}
}

Then, the second equation in~\eqref{eq:General} leads to
\EQ{
\Upsilon_\pm(z) = \frac{1}2\left(\frac{\Phi_\pm}{\Phi_\mp}-1\right)= \mp\frac{1}{z\pm1},
\label{eq:SolSpectators2}
}
which is independent of $\lambda$. But then, as anticipated, the first equation in~\eqref{eq:General} requires that $\lambda=1$. 

We therefore find that if we do not make any assumption on the algebraic structure of $[\mathfrak{f}^\perp, \mathfrak{f}^\perp]$, then we can construct a Lax connection only for $\lambda=1$. In particular, we find
\EQ{
\AA_\pm (z)\big|_{\lambda=1}&= \frac{z}{z\pm1}\, \left(A_\pm + j^\mathfrak{f}_\pm \right) -j^\mathfrak{f}_\pm\mp\frac{1}{z\pm1}\, j^\perp_\pm\\[5pt]
&=\frac{z}{z\pm1}\, A_\pm  -\frac{1}{1\pm z}\, j_\pm.
\label{eq:Lax-NATD}
}
This is in fact the Lax connection for the NATD model where a subgroup $F\subset G$ is dualised~\cite{Borsato:2016pas}.

\subsection{$\left[\mathfrak{f}^\perp,\mathfrak{f}^\perp\right]\subset \mathfrak{f}^\perp$: the trivial case of direct product form}
A second possibility would be
\EQ{
\left[\mathfrak{f}^\perp,\mathfrak{f}^\perp\right]\subset \mathfrak{f}^\perp,
}
which means that $\mathfrak{f}^\perp$ is a subalgebra of~$\mathfrak{g}$. 
Actually, since $\left[\mathfrak{f},\mathfrak{f}^\perp\right]\subset \mathfrak{f}^\perp$, it also satisfies that 
$\left[\mathfrak{g},\mathfrak{f}^\perp\right]\subset \mathfrak{f}^\perp$ and, thus, $\mathfrak{f}^\perp$ is an ideal. 
Since the bilinear form is non-degenerate, $\mathfrak{f}\cap \mathfrak{f}^\perp=\{0\}$ (as vector spaces) and one can check that $[\mathfrak{f}, \mathfrak{f}^\perp]=\{0\}$.\footnote{Let $X\in \mathfrak{f}$, $Y^\perp\in\mathfrak{f}^\perp$, and $Z\in\mathfrak{g}$. Using that the bilinear form is ad-invariant and that $\mathfrak{f}^\perp$ is an ideal,
$
\Tr\left\{\left[X,Y^\perp\right]\, Z\right\} =\Tr\left\{X\left[Y^\perp,Z\right]\right\} =0,\quad \forall\, Z\in \mathfrak{g}\;\; \Rightarrow\;\; \left[X,Y^\perp\right]=0.
$
}
The conclusion is that, in this case, $\mathfrak{g}$ is the direct sum of the two Lie subalgebras $\mathfrak{f}$ and $\mathfrak{f}^\perp$. Correspondingly,
\EQ{
G=F\times F^\perp \;\; \Rightarrow\;\; F\backslash G \simeq F^\perp
}
and the coset space is the group manifold corresponding to the subalgebra $\mathfrak{f}^\perp$. 
Importantly, because now $\mathfrak{f}$ and $\mathfrak{f}^\perp$ commute, we do not have to impose the equations in the third row of table~\ref{tab:int} and $\alpha$ remains \emph{generic}, without the need to fix it to the value in~\eqref{eq:alpha}.

Then, assuming that $\mathfrak{f}^\perp$ is non-abelian, we only have to solve the equations on the first and fourth row of table~\ref{tab:int}. 
They can be solved as
\EQ{
\Upsilon_\pm= -\frac{\mu}{\mu\pm1},
}
where $\mu$ is a second spectral parameter independent of $z$. The fact that we can take two independent spectral parameters is related to the direct product form of $G$.\footnote{Obviously, other parameterisations are also possible, for example $\Upsilon_\pm(\mu) = \mp\frac{1}{\mu\pm1}$.}
This gives rise to the Lax connection
\EQ{
\AA_\pm (z,\mu)= \frac{z}{z\pm 1}\, \frac{2}{1+\lambda}\, \left(A_\pm + j^\mathfrak{f}_\pm \right) -j^\mathfrak{f}_\pm -\frac{\mu}{\mu\pm1}\, j^\perp_\pm,
\label{eq:PossibilityTwo}
}
where we have to remind that $\alpha$ remains arbitrary. This Lax connection coincides with~\eqref{eq:Lax-NATD} when $\lambda=1$ and $\mu=\nicefrac1z$.
As discussed in section~\ref{sec:action}, after gauge fixing one can take $g\in F\backslash G$, so that in this case $g\in F\backslash G=F^\perp$ and $j^\mathfrak{f}_\pm =0$, resulting in
\EQ{
\AA_\pm (z)= \frac{z}{z\pm 1}\, \frac{2}{1+\lambda}\, A_\pm    -\frac{\mu}{\mu\pm1}\, j^\perp_\pm.
\label{eq:PossibilityTwo-gf}
}
The two contributions are respectively the Lax connection of the usual $\lambda$-deformation of the PCM corresponding to $F$ with spectral parameter $z$, and the Lax connection of the undeformed PCM corresponding to $F^\perp$ with spectral parameter $\mu$. Since $[\mathfrak{f},\mathfrak{f}^\perp]=\{0\}$ they give rise to decoupled zero-curvature conditions. Note also that in~\eqref{eq:Action} the coupling constant of the PCM corresponding to $F^\perp$ is $\alpha\kappa^2$. Being $\alpha$  arbitrary, $\alpha\kappa^2$ is effectively an independent coupling constant for that PCM.

All this shows that this case is not interesting and we will ignore it: the deformation in $F$ is just the standard $\lambda$-deformation and, since $G=F\times F^\perp$ is a direct product, the spectators, which are the degrees of freedom associated to $\mathfrak{f}^\perp$, are not involved in a non-trivial way in the construction.

\subsection{$\left[\mathfrak{f}^\perp,\mathfrak{f}^\perp\right]\subset \mathfrak{f}$: $\mathbb Z_2$ structure and integrability} 
The third and most interesting case is identified by making the assumption
\EQ{
\left[\mathfrak{f}^\perp,\mathfrak{f}^\perp\right]\subset \mathfrak{f}.
}
Since $\left[\mathfrak{f},\mathfrak{f}\right]\subset \mathfrak{f}$ and $\left[\mathfrak{f},\mathfrak{f}^\perp\right]\subset \mathfrak{f}^\perp$, it means that $F\backslash G$ is a symmetric space. Then, $\left[ j_+^\perp, j_-^\perp\right]^\perp=0$  {so that we don't have to impose the equation in the last line of table~\ref{tab:int}. We do have to worry about the second line (i.e. eq.~\eqref{eq:mcH})} which becomes
\EQ{
\Upsilon_+^2 = -\frac{\Phi_+}{\Phi_-}\, \Psi_+\Psi_-=\frac{\Phi_+}{\Phi_-}\, \left(\lambda\Phi_+\Phi_- -1\right).
}
Using~\eqref{eq:standard}, there are two solutions
\EQ{
\Upsilon_\pm(z)=\sigma\,\frac{\sqrt{(1+\lambda)^2 -(1-\lambda)^2\, z^2}}{(1\pm z)\,(1+\lambda)},
\label{eq:SolSpectators1}
}
with $\sigma=+1$ or $-1$. 
We therefore find the following Lax connection
\EQ{
\AA_\pm (z)= \frac{z}{z\pm 1}\, \frac{2}{1+\lambda}\, \left(A_\pm +j^\mathfrak{f}_\pm\right) -j^\mathfrak{f}_\pm + \sigma\,\frac{\sqrt{(1+\lambda)^2 -(1-\lambda)^2\, z^2}}{(1\pm z)\,(1+\lambda)}\;  j^\perp_\pm .
\label{eq:SolSpectators1b}
}
It is worth noticing that $\sigma\to-\sigma$ is equivalent to the $\mathbb{Z}_2$ symmetry $\mathfrak{f}^\perp \to -\mathfrak{f}^\perp$ of the symmetric space. Therefore, the choice of the value of $\sigma$ is just a matter of convention, and we will choose $\sigma=-1$ so that $\AA_\pm(z)$ coincides with~\eqref{eq:Lax-NATD} at $\lambda=1$.

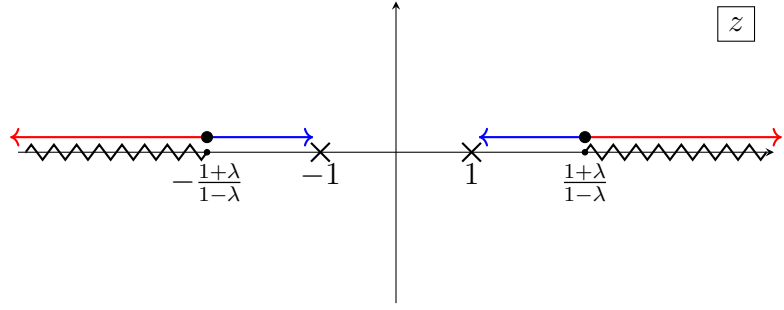
\begin{figure}
    \centering
  \begin{tikzpicture}
  \tikzset{
    cross/.pic = {
    \draw[rotate = 45] (-#1,0) -- (#1,0);
    \draw[rotate = 45] (0,-#1) -- (0, #1);
    }
}
    \draw [-stealth](-5,0) -- (5,0);
    \draw [-stealth](0,-2) -- (0,2);
         \draw (4.5,1.7) node[shape=rectangle,draw]{$z$};
     \draw[thick,blue,
        decoration={markings, mark=at position 1 with {\arrow{>}}},
        postaction={decorate}
        ] (-2.5,0.2) -- (-1.1,0.2) ;
  \draw[thick,blue,
        decoration={markings, mark=at position 1 with {\arrow{>}}},
        postaction={decorate}
        ] (2.5,0.2) -- (1.1,0.2);
             \draw[thick,red,
        decoration={markings, mark=at position 1 with {\arrow{>}}},
        postaction={decorate}
        ] (-2.5,0.2) -- (-5.1,0.2) ;
             \draw[thick,red,
        decoration={markings, mark=at position 1 with {\arrow{>}}},
        postaction={decorate}
        ] (2.5,0.2) -- (5.1,0.2) ;
        \draw[thick,decoration = {zigzag,segment length = 3mm, amplitude = 1mm},decorate] (2.5,0)--(4.9,0);
        \draw[thick,decoration = {zigzag,segment length = 3mm, amplitude = 1mm},decorate] (-2.5,0)--(-4.9,0);
    \draw[black,fill=black] (2.5,0) circle (0.2ex) node[below] {$\frac{1+\lambda}{1-\lambda}$};
    \draw[black,fill=black] (-2.5,0) circle (0.2ex) node[below] {$-\frac{1+\lambda}{1-\lambda}$};
    \draw[black,fill=black] (2.5,0.2) circle (0.4ex) node[] {};
    \draw[black,fill=black] (-2.5,0.2) circle (0.4ex) node[] {};
    \path (1,0) pic[thick] {cross=5pt} node[below, black] {$1$};
        \path (-1,0) pic[thick] {cross=5pt}  node[below, black] {$-1$};
\end{tikzpicture}
    \caption{Analytic structure in one Riemann sheet parameterised by $z$ and the behaviour in the $\lambda\to 0$ (blue) and $\lambda\to 1$ (red) limits. The crosses at $z=\pm 1$ mark the poles, while the zigzag lines are the branch cuts. In the $\lambda\to 0$ limit the position of the branch points coincide with that of the poles, while in the $\lambda\to 1$ limit the branch points go to infinity.}
    \label{fig:z-sheet}
\end{figure}

The analytic structure of the Lax connection~\eqref{eq:SolSpectators1b} as a function of the complex variable $z$ depends on the value of $\lambda$.
At $\lambda=1$, $\AA_\pm (z)$ is the Lax connection for the NATD model where a subgroup $F\subset G$ is dualised~\cite{Borsato:2016pas}, and it only exhibits two simple poles at $z=\pm1$. However, when $0<\lambda<1$ it has two square-root branch points at
\EQ{
z=\pm \frac{1+\lambda}{1-\lambda}=\pm z_B(\lambda)\,,
}
in addition to the poles.
In this case we can make sense of the Lax connection in terms of a 2-cover of the complex $z$-plane with cuts along $(-\infty,-z_B(\lambda))$ and $(+z_B(\lambda),+\infty)$. Note that $z_B(\lambda)\to \infty$ when $\lambda\to1$, which agrees with the fact that the cuts dissappear in this limit. The position of the poles and branch cuts in the $z$-plane and their dependence on $\lambda$ is summarised in fig.~\ref{fig:z-sheet}.

At $\lambda=0$ the Lax connection becomes
\EQ{
\AA_\pm (z)\big|_{\lambda=0}= \frac{2z}{z\pm 1}\, \left(A_\pm +j^\mathfrak{f}_\pm\right) -j^\mathfrak{f}_\pm -\sqrt{\frac{1\mp z}{1\pm z}}\;  j^\perp_\pm.
\label{NewLax}
}
In the naive $\lambda\to0$ limit where all the fields are independent of $\lambda$, eq.~\eqref{eq:eomANew} implies that $A_\pm +j^\mathfrak{f}_\pm=
{\cal O}(\lambda)$ and, hence,
\EQ{
\AA_\pm (z)\big|_{\lambda=0}=-j^\mathfrak{f}_\pm -\sqrt{\frac{1\mp z}{1\pm z}}\;  j^\perp_\pm.
}
This is just the Lax connection of the $F\backslash G$ symmetric space coset sigma model 
\EQ{
\AA_\pm (z)\big|_{\lambda=0}=-j^\mathfrak{f}_\pm - \mu^{\pm1}  j^\perp_\pm, \qquad \mu= \sqrt{\frac{1- z}{1+ z}}:=\mu(z)\,,
} 
with spectral parameter $\mu$.\footnote{In terms of $j_\pm$, the Maurer--Cartan identity reads $[\partial_+- j_+, \partial_- -j_-]=0$, which explains the minus sign in front of $j^\mathfrak{f}_\pm$ and $j^\perp_\pm$.} Note that $\mu(-1)=\infty$ while $\mu(1)=0$.
In section~\ref{sec:lim} it was shown that this naive limit corresponds to keeping the PCM coupling $\kappa^2$ fixed while sending $k\to0$, so that we actually recover the action of the symmetric space coset sigma model~\eqref{eq:naive-lambda0}. But, consequently, only $\text{dim}(G)-\text{dim}(F)$ degrees of freedom survive.

In section~\ref{sec:lim} it was also argued that to keep all the $\text{dim}(G)$ degrees of freedom we have to take  the correlated limit
\EQ{
\lambda\to0, \qquad j_\pm\to \sqrt{\lambda}\; \partial_\pm \phi,
}
which corresponds to keeping the gWZW coupling~$k$ fixed while sending $\kappa^2\to\infty$. In this limit the action becomes~\eqref{eq:lambda-to-zero-1}, which is the sum of a WZW term associated to $F$ and a decoupled free boson that takes values in $\mathfrak{f}^\perp$. However, since eq.~\eqref{eq:eomANew} also implies that $j_\pm^\mathfrak{f} + A_\pm = \cal O(\lambda)$ in this limit,
the Lax connection~\eqref{eq:SolSpectators1b} vanishes at $\lambda=0$ {(always up to a gauge transformation).}
Strictly speaking, this means that the deformed theory at $\lambda=0$ does not admit the usual  integrability description, because its equations of motion cannot be written in terms of the Lax connection. Nevertheless, it is remarkable that we can recover the equations of motion by looking at the first orders in the expansion of the zero curvature equations around $\lambda=0$. 
To spell this out, let us write the correlated limit as follows
\EQ{
\lambda\to0, \qquad j_\pm\to \sqrt{\lambda}\; \partial_\pm \phi
+ \lambda \delta_\pm +
{\cal O}(\lambda^{\nicefrac32})\,,
}
to take into account the effect of additional terms of higher order in the expansion around $\lambda=0$. Then, eq.~\eqref{eq:eomANew} implies that
\EQ{
&
j_+^\mathfrak{f} + A_+ = \lambda\, \CF^{-1}\partial_+\CF + {\cal O}(\lambda^{\nicefrac32}),\\[5pt]
&
j_-^\mathfrak{f} + A_- = -\lambda\, \partial_-\CF\CF^{-1} + {\cal O}(\lambda^{\nicefrac32}),
}
which leads to the following expansion of the Lax connection~\eqref{eq:SolSpectators1b} around $\lambda=0$:
\EQ{
&
\AA_+(z) = \sqrt{\lambda} \left(-\partial_+\phi^f -\mu(z)\, \partial_+\phi^\perp\right)+ \\[5pt]
&\qquad\qquad
+\lambda \left(\frac{2z}{z+1}\, \CF^{-1}\partial_+\CF - \delta_+^\mathfrak{f} - \mu(z)\, \delta_+^\perp\right) + {\cal O}(\lambda^{\nicefrac32}),\\[5pt]
&
\AA_-(z) = \sqrt{\lambda} \left(-\partial_-\phi^f -\mu^{-1}(z)\, \partial_-\phi^\perp\right) + \\[5pt]
&\qquad\qquad
+\lambda \left(-\frac{2z}{z-1}\,\partial_-\CF \CF^{-1}- \delta_-^\mathfrak{f} - \mu^{-1}(z)\, \delta_-^\perp\right) + {\cal O}(\lambda^{\nicefrac32}).
}
Then, the leading order term of the zero-curvature equation reads
\EQ{
0=[\partial_++ \AA_+(z), \partial_- +\AA_-(z)]=\sqrt{\lambda}\left(\mu(z)-\mu^{-1}(z)\right)\, \partial_+\partial_- \phi^\perp + {\cal O}(\lambda).
}
Since it has to vanish for any value of $z$, it provides the equations of motion of the free boson in~\eqref{eq:lambda-to-zero-1}; namely, $\partial_+\partial_- \phi^\perp=0$. At $\mathcal O(\lambda)$ we find
\EQ{
&
-\frac{2z}{z-1}\,\partial_+\left(\partial_-\CF \CF^{-1}\right)- \frac{2z}{z+1}\, \partial_-\left(\CF^{-1}\partial_+\CF\right)\\[5pt]
&\qquad
+\left[\partial_+\phi^\mathfrak{f}, \partial_-\phi^\mathfrak{f}\right]+\left[\partial_+\phi^\perp, \partial_-\phi^\perp\right]-\partial_+\delta_-^\mathfrak{f} +\partial_-\delta_+^\mathfrak{f}\\[5pt]
&\qquad
+\mu(z)\left(\left[\partial_+\phi^\perp, \partial_-\phi^\mathfrak{f}\right]+\partial_-\delta_+^\perp\right)
+\mu^{-1}(z)\left(\left[\partial_+\phi^\mathfrak{f}, \partial_-\phi^\perp \right]-\partial_+\delta_-^\perp\right),
}
which also have to vanish for any value of $z$.
Therefore, the residues at the simple poles at $z=\pm1$ provide the equations of motion of the WZW action in~\eqref{eq:lambda-to-zero-1}:
\EQ{
\partial_-(\CF^{-1}\partial_+ \CF)=\partial_+(\partial_-\CF \CF^{-1})=0.
}
The rest of the terms provide equations that allow one to fix the form of the higher order terms $\delta_\pm$ as a function of $\phi$ and 
$\partial_\pm \phi$.

\subsubsection{Alternative parameterisations}
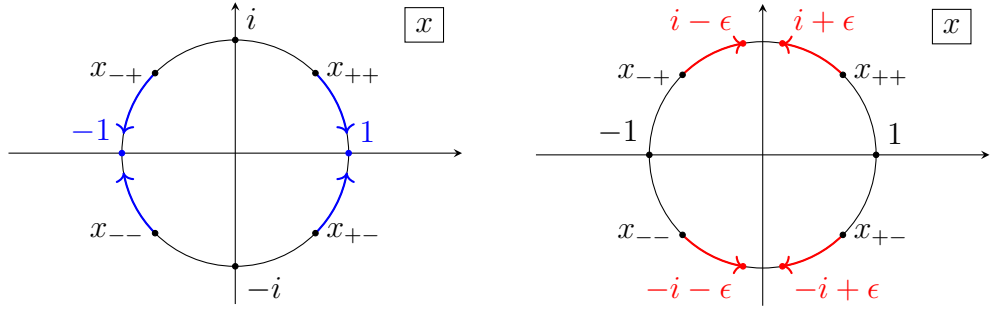
\begin{figure}
    \centering
  \begin{tikzpicture}
    \draw [-stealth](-3,0) -- (3,0);
    \draw [-stealth](0,-2) -- (0,2);
         \draw (2.5,1.7) node[shape=rectangle,draw]{$x$};
    \draw[black] (0,0) circle (1.5) ;
     \draw[thick,blue,
        decoration={markings, mark=at position 1 with {\arrow{>}}},
        postaction={decorate}
        ] (1.5/1.41421,1.5/1.41421) arc
    [
        start angle=45,
        end angle=10,
        radius=1.5
    ] ;
         \draw[thick,blue,
        decoration={markings, mark=at position 1 with {\arrow{>}}},
        postaction={decorate}
        ] (1.5/1.41421,-1.5/1.41421) arc
    [
        start angle=-45,
        end angle=-10,
        radius=1.5
    ] ;
         \draw[thick,blue,
        decoration={markings, mark=at position 1 with {\arrow{>}}},
        postaction={decorate}
        ] (-1.5/1.41421,1.5/1.41421) arc
    [
        start angle=135,
        end angle=170,
        radius=1.5
    ] ;
         \draw[thick,blue,
        decoration={markings, mark=at position 1 with {\arrow{>}}},
        postaction={decorate}
        ] (-1.5/1.41421,-1.5/1.41421) arc
    [
        start angle=225,
        end angle=190,
        radius=1.5
    ] ;
    \draw[black,fill=black] (1.5/1.41421,1.5/1.41421) circle (0.2ex) node[right] {$x_{++}$};
    \draw[black,fill=black] (1.5/1.41421,-1.5/1.41421) circle (0.2ex) node[right] {$x_{+-}$};
    \draw[black,fill=black] (-1.5/1.41421,1.5/1.41421) circle (0.2ex) node[left] {$x_{-+}$};
    \draw[black,fill=black] (-1.5/1.41421,-1.5/1.41421) circle (0.2ex) node[left] {$x_{--}$};
    \draw[blue,fill=blue] (1.5,0) circle (0.2ex) node[above right] {$1$};
    \draw[blue,fill=blue] (-1.5,0) circle (0.2ex) node[above left] {$-1$};
    \draw[black,fill=black] (0,1.5) circle (0.2ex) node[above right] {$i$};
    \draw[black,fill=black] (0,-1.5) circle (0.2ex) node[below right] {$-i$};
\end{tikzpicture}
\qquad
      \begin{tikzpicture}
    \draw [-stealth](-3,0) -- (3,0);
    \draw [-stealth](0,-2) -- (0,2);
    \draw (2.5,1.7) node[shape=rectangle,draw]{$x$};
    \draw[black] (0,0) circle (1.5) ;
     \draw[thick,red,
        decoration={markings, mark=at position 1 with {\arrow{>}}},
        postaction={decorate}
        ] (1.5/1.41421,1.5/1.41421) arc
    [
        start angle=45,
        end angle=80,
        radius=1.5
    ] ;
         \draw[thick,red,
        decoration={markings, mark=at position 1 with {\arrow{>}}},
        postaction={decorate}
        ] (1.5/1.41421,-1.5/1.41421) arc
    [
        start angle=-45,
        end angle=-80,
        radius=1.5
    ] ;
         \draw[thick,red,
        decoration={markings, mark=at position 1 with {\arrow{>}}},
        postaction={decorate}
        ] (-1.5/1.41421,1.5/1.41421) arc
    [
        start angle=135,
        end angle=100,
        radius=1.5
    ] ;
         \draw[thick,red,
        decoration={markings, mark=at position 1 with {\arrow{>}}},
        postaction={decorate}
        ] (-1.5/1.41421,-1.5/1.41421) arc
    [
        start angle=225,
        end angle=260,
        radius=1.5
    ] ;
    \draw[black,fill=black] (1.5/1.41421,1.5/1.41421) circle (0.2ex) node[right] {$x_{++}$};
    \draw[black,fill=black] (1.5/1.41421,-1.5/1.41421) circle (0.2ex) node[right] {$x_{+-}$};
    \draw[black,fill=black] (-1.5/1.41421,1.5/1.41421) circle (0.2ex) node[left] {$x_{-+}$};
    \draw[black,fill=black] (-1.5/1.41421,-1.5/1.41421) circle (0.2ex) node[left] {$x_{--}$};
    \draw[black,fill=black] (1.5,0) circle (0.2ex) node[above right] {$1$};
    \draw[black,fill=black] (-1.5,0) circle (0.2ex) node[above left] {$-1$};
    \draw[red,fill=red] (0.260472,1.47721) circle (0.2ex) node[above right] {$i+\epsilon$};
        \draw[red,fill=red] (-0.260472,1.47721) circle (0.2ex) node[above left] {$i-\epsilon$};
    \draw[red,fill=red] (0.260472,-1.47721) circle (0.2ex) node[below right] {$-i+\epsilon$};
        \draw[red,fill=red] (-0.260472,-1.47721) circle (0.2ex) node[below left] {$-i-\epsilon$};
\end{tikzpicture}
    \caption{Limiting values of the poles of the Lax connection when $\lambda\to 0$ (left, blue) and $\lambda\to 1$ (right, red).}
    \label{fig:poles-x}
\end{figure}

We can explicitly describe the 2-cover of the $z$-plane that arises when $\lambda<1$ simply by changing the spectral parameter with the redefinition
\EQ{
z=\frac{1+\lambda}{1-\lambda}\, \cos(\theta),
\label{eq:ZmapTheta}
}
which leads to
\EQ{
\Upsilon_\pm(\theta)=-\frac{(1-\lambda)\sin(\theta)}{(1-\lambda)\pm(1+\lambda)\cos(\theta)}.
}
In this new parameterisation, the Lax connection  looks like
\EQ{
\AA_\pm (\theta)&= \frac{2\cos(\theta)}{(1+\lambda)\cos(\theta)\pm (1-\lambda)}\, \left(A_\pm +   j^\mathfrak{f}_\pm \right)- j^\mathfrak{f}_\pm \\[5pt]
&
-\frac{(1-\lambda)\sin(\theta)}{(1-\lambda)\pm (1+\lambda)\cos(\theta)}\;  j^\perp_\pm,
\label{eq:LaxResult-theta}
}
with no explicit square root.
From~\eqref{eq:ZmapTheta} we see that the two Riemann sheets parameterised by $z$ are related by $\theta\to -\theta$.
An alternative description is provided by the Zhukovsky-like transformation\footnote{As before, the parameterisation is motivated by the resolution of the square root, that in this case reads $\sqrt{(\lambda+1)^2 -(\lambda-1)^2\, z^2}=\frac{i}2\, (\lambda+1) \left(x-\nicefrac1x\right)$.}
\EQ{
z=\frac{1}2 \, \frac{1+\lambda}{1-\lambda}\, \left(x+\frac{1}{x}\right):= z(x,\lambda),
\label{eq:ZmapX}
}
which is equivalent to~\eqref{eq:ZmapTheta} by taking $x=e^{-i\theta}$. 
The inverse of~\eqref{eq:ZmapX} is 
\EQ{
x=\frac{1-\lambda}{1+\lambda}\, z \pm \sqrt{\left(\frac{1-\lambda}{1+\lambda}\right)^2 z^2-1}\,:=\, x_\pm(z,\lambda)= \frac{1}{x_\mp(z,\lambda)},
\label{eq:XmapZ}
}
which shows that each value of $z$ has images $x$ and $\nicefrac1x$ on the two branches of the cover when $\lambda\not=1$.
In particular, the branch points at $z=\pm z_B(\lambda)$ correspond to $x=\pm1$ (or $\theta=0,\pi$). The two poles at $z=s=\pm1$ give rise to poles located at
\EQ{
x_\pm(s,\lambda)=\frac{(1-\lambda)s \pm 2i\sqrt{\lambda}}{1+\lambda}\equiv x_{s\pm}(\lambda), \qquad |x_{s\pm}(\lambda)|=1,
}
so that
\EQ{
z\pm 1=\frac{1}{2x}\,\frac{1+\lambda}{1-\lambda}\, \left(x- x_{\mp-}(\lambda)\right)\left(x-x_{\mp+}(\lambda)\right).
}
In terms of the new spectral parameter~$x$, the Lax connection reads
\EQ{
\widetilde{\AA}_\pm(x;\lambda)&=\frac{2(x^2+1)}{\left(x-x_{\mp-}(\lambda)\right)\, \left(x-x_{\mp+}(\lambda)\right)}\, \left(A_\pm + j_\pm^\mathfrak{f}\right)-j_\pm^\mathfrak{f}\\[5pt]
&
\mp i\, \frac{(1-\lambda)\, (x^2-1)}{(1+\lambda)\, \left(x-x_{\mp-}(\lambda)\right)\, \left(x-x_{\mp+}(\lambda)\right)}\; j_\pm^\perp,
\label{eq:LaxResult-x}
}
which exhibits a total of 4 simple poles in the $x$ plane, all located on the unit circle $|x|=1$, see figure~\ref{fig:poles-x}.

It is important to emphasise that the transformation $z=z(x,\lambda)$ is not defined at $\lambda=1$.  In fact, in the  limit $\lambda\to 1$ the poles behave as
\EQ{
z=s=\pm1 \;\; \Rightarrow\;\; x_{s\pm}(\lambda)\to \pm i +\frac{1-\lambda}2\,s,
}
so that in each branch the two poles corresponding to $z=+1$ and $z=-1$ go to a single point $x= \pm i$.\footnote{Note that $x_\pm(z,\lambda=1)= \pm i$ for any value of~$z$, which exhibits that the transformation is actually not defined at $\lambda=1$.} The correct description of the $\lambda\to1$ limit is provided by the relationship between the Lax connections~\eqref{eq:SolSpectators1b} and~\eqref{eq:LaxResult-x}, which reads
\EQ{
\widetilde{\AA}_\pm(x(z,\lambda);\lambda)=\AA_\pm(z;\lambda), \qquad \forall\;\lambda\not=1
}
for either $x=x_+(z,\lambda)$ or $x=x_-(z,\lambda)$.  
It shows that the correct form of the limit is
\EQ{
x\sim \pm i+(1-\lambda)\frac{z}{2}.
}
In other words, we should view the $\lambda\to 1$ limit as a ``zoom-in limit'' around $x=i$ or $x=-i$, depending on the branch that is selected. See also figure~\ref{fig:poles-x}.

When $\lambda\to 0$, the 4 simple poles collapse to just 2 double poles at
\EQ{
x_{+\pm}\to 1,\quad\text{and}\quad x_{-\pm}\to -1, 
}
see also figure~\ref{fig:poles-x}. However, in the correlated limit with $j\approx\mathcal O(\sqrt{\lambda})$, the Lax connection vanishes at $\lambda=0$.
We refer to the discussion in the first part of this section for more details.

In summary, one recovers two simple poles in the limit $\lambda\to 1$ instead of four poles, as expected from the interpretation of the corresponding $\sigma$-model. This is a consequence of the ``zoom-in limit'' around either $x=+i$ or $x=-i$ retaining only one of the two branches. Moreover, the Lax connection vanishes in the strict $\lambda\to 0$ limit, and the four single poles emerge from the two double poles at $x=\pm1$ when $\lambda$ is switched on. We see that the double cover of $z$ is necessary to have an interpolation between the two descriptions at $\lambda=0$ and $\lambda=1$.

\subsection{$F$ abelian}
Finally, let us comment on the special case of $F$ abelian. In this case, we don't have to impose the equations in the first line of table~\ref{tab:int}. That means that the parameter $\alpha$  remains generic and we don't need to fix it in terms of $\lambda$.  It follows that the Lax connection depends only on $\alpha$ (not on $\lambda$), and it may be obtained from the previous formulas (for example~\eqref{eq:SolSpectators1b}) by implementing the formal substitution $\lambda = 2\alpha-1$. The rest of the analysis is similar to the discussion above: also in this case, if we do not consider the symmetric space case and allow for the fourth line in  table~\ref{tab:int}, the second and fourth lines are incompatible unless $\alpha=1$. We stress that also in the abelian case the deformation is still non-trivial: integrability still demands that $F\backslash G$ must be symmetric, and the Lax connection will be given by~\eqref{eq:SolSpectators1b}, for example.

Notice that the fact that the equations of motion are $\lambda$-independent is true only because, to construct the Lax, we  write them in terms of $A_\pm$ and the projections of $j_\pm$. In particular, the $\lambda$-dependence will reappear when writing $A_\pm$ as in~\eqref{eq:EqsGaugeFields}. To be concrete, if we parameterise $\CF = \exp(\beta \varphi)$, where $\varphi\in\mathfrak{f}$ is a field and $\beta\in \mathbb R$ is a normalisation parameter, then 
\EQ{
A_\pm= \pm \frac{\lambda\beta}{1-\lambda}\partial_\pm\varphi-j_\pm^{\mathfrak{f}}.
\label{eq:A-ab}
} 
It is actually interesting to look at how the action~\eqref{eq:S-lambda-sub} simplifies when $F$ is abelian. Up to boundary terms, we find
\EQ{
&
S[g,\varphi;\alpha]=-\frac{k}{\pi\lambda}\int d^2x\, \Tr\left\{ \alpha(1-\lambda)\, j_+^\perp j_-^\perp  +\frac{\lambda\beta^2}{2}\frac{1+\lambda}{1-\lambda} \partial_+\varphi\partial_-\varphi  \right.\\[5pt]
&\qquad \qquad\qquad\qquad \qquad\qquad
\left.-\lambda\beta\ \varphi(\partial_+j_-^{\mathfrak{f}}-\partial_-j_+^{\mathfrak{f}})\right\}.
}
{In the limit $\lambda\to 0$, if we take $j_\pm\approx \sqrt{\lambda} \partial_\pm \phi$ as in~\eqref{eq:lambda-to-zero-1}, we can fix for example the normalisation $\beta=1$ and obtain
\EQ{
&
S[g,\varphi;\alpha]=-\frac{k}{\pi}\int d^2x\, \Tr\left\{ \alpha^*\, \partial_+\phi^\perp \partial_-\phi^\perp  +\frac{1}{2} \partial_+\varphi\partial_-\varphi  \right\},\label{Eq:action.Abelian.zero-1}
}
so that we have just a collection of dim$(G)$ free bosons.
If, instead, we take the more general~\eqref{eq:lambdaZeroLimit-2} and~\eqref{eq:lambda-to-zero-2} with the parameterisation $U=e^{u}$ and $\beta=1$, we  find
\EQ{
&
S[g,\varphi;\alpha]=-\frac{k}{\pi}\int d^2x\, \Tr\left\{ \alpha^*\, \partial_+\phi^\perp \partial_-\phi^\perp  +\frac{1}{2} \partial_+\varphi\partial_-\varphi   +\partial_+\varphi\partial_-u-\partial_+u\partial_-\varphi)\right\}.\label{Eq:action.Abelian.zero-2}
}
We see that the degrees of freedom in $u$ are pure gauge, because the last two terms combine, in fact, in a total derivative, and after dropping them we go back to~\eqref{Eq:action.Abelian.zero-1}.
}
In the limit $\lambda\to 1$ with $k(1-\lambda)=\kappa^2$ fixed, instead, it is convenient to take $\beta=1-\lambda$ and we obtain
\EQ{
&
S[g,\varphi;\alpha]=-\frac{\kappa^2}{\pi}\int d^2x\, \Tr\left\{ \alpha\, j^\perp_+j^\perp_-  +\partial_+\varphi\partial_-\varphi - \varphi(\partial_+j_-^{\mathfrak{f}}-\partial_-j_+^{\mathfrak{f}})\right\}.\label{Eq:action.Abelian.one}
}
It is easy to check that this result is reproduced by using~\eqref{eq:A-ab} in~\eqref{eq:NATDaction1} with the identifications $\varphi=\nu$ and $\alpha=1$. Therefore,  the interpretation of the $\lambda\to 1$ limit is abelian T-duality as expected.

\section{An example}\label{sec:ex}

Below we consider the simple case where $G=SU(2)$ and the subgroup is abelian, i.e. $F=U(1)$. Choosing a particular gauge slice, we parameterise the $SU(2)/U(1)$ coset representative $q$ and the group element for the $U(1)$ subgroup $\CF$ respectively as
\EQ{q=e^{\frac{i\theta}{2}\sigma_2}
e^{\frac{i\phi}{2}\sigma_3}\,,\quad
\CF=e^{\frac{i\varphi}{2}\sigma_3} \, .
}
Then, the action \eqref{eq:S-lambda-sub} leads to the following $\sigma$-model
\EQ{
\label{action.Abelian.example}
\begin{split}
S=\frac{k}{2\pi}\int d^2x&\left\{\frac{\alpha(1-\lambda)}{\lambda}\left(\partial_+\theta\partial_-\theta+
\sin^2\theta \partial_+\phi\partial_-\phi\right)+\frac{1+\lambda}{2(1-\lambda)}\partial_+\varphi\partial_-\varphi\right.\\
&\left.-\cos\theta(\partial_+\phi\partial_-\varphi-\partial_+\varphi\partial_-\phi)\right\} \,,
\end{split}}
which is integrable for every value of the two parameters $\lambda$ and $\alpha$. Note that the
parameter $\alpha$ is a second genuine parameter, since the subgroup $F=U(1)$ is abelian.
Equivalently, the action \eqref{action.Abelian.example} describes the abelian T-dual of the PCM on the squashed 3-sphere fibered over the 2-sphere, that is 
\EQ{\label{Abelian-T-dual}S=\frac{k}{\pi}\frac{1-\lambda}{1+\lambda}\int d^2x&\Big(
(\partial_+\tilde\varphi+\cos\theta\partial_+\phi)(\partial_-\tilde\varphi+\cos\theta\partial_-\phi)\\
&+\frac{\alpha(1+\lambda)}{2\lambda}\left(\partial_+\theta\partial_-\theta+
\sin^2\theta \partial_+\phi\partial_-\phi\right) \Big)\,,}
where $\tilde\varphi$ is the T-dual variable of $\varphi$. The latter in an integrable model and it corresponds to the trigonometric deformation of the PCM on $S^3$ of~\cite{Cherednik:1981df}\footnote{For $\alpha=\frac{2\lambda}{1+\lambda}$ the target space of \eqref{Abelian-T-dual} is the round $S^3$.\label{foot.round}}, or equivalently the $SU(2)$ Yang--Baxter $\sigma$-model  of~\cite{Klimcik:2002zj},\footnote{We thank Ben Hoare for raising this point.}
ensuring the integrability of the action \eqref{action.Abelian.example}, since
 the abelian T-duality is a canonical transformation~\cite{Alvarez:1994wj}.
 
The target space of the $\sigma$-model  \eqref{action.Abelian.example} is compact 
and non-singular for $\lambda\in(0,1)$ and $\alpha>0$. Moreover, the above action is invariant under the symmetry \eqref{eq:sym-alpha0} and \eqref{eq:sym-alpha}, where $\CF\to \CF^{-1}$ is implemented simply by $\varphi\to-\varphi$. Additionally, in this case  from \eqref{ScalarField} one obtains a constant scalar dilaton
\EQ{\Phi = - \frac{1}{2} \log \Big( k \frac{1 - \lambda}{\lambda} \Big) \, .}
To take the $\lambda \to 0$ limit of \eqref{action.Abelian.example}\footnote{Assuming 
$\displaystyle\alpha^*=\lim_{\lambda\to 0}\alpha\neq0$, along the discussion in section \ref{sec:lim}.}, we need to rescale the coset fields with $\lambda$ around some values as
\EQ{
\theta\to \frac{\pi}{2}+\sqrt{\lambda}\theta+\ldots,
\qquad\qquad
\phi\to\sqrt{\lambda}\phi+\ldots,
}
where the dots denote subleading orders in $\lambda$ (actually, we could do the expansion around any non-vanishing constant value $\theta_0$ not necessarily equal to $\nicefrac{\pi}{2}$). Comparing to~\eqref{eq:lambdaZeroLimit} this means that we are taking the expansion of $q$ around a non-trivial constant value $q_0$
\EQ{
q\approx\left(1+\sqrt{\lambda}\left(\frac{i\theta}{2}\sigma_2-\frac{i\phi}{2}\sigma_1\right)+\ldots\right)q_0,
\qquad\qquad
q_0=e^{\frac{i\pi}{4}\sigma_2}.
}
Then the action is
\EQ{S=\frac{k}{2\pi}\int d^2x\left\{\alpha(\partial_+\theta\partial_-\theta+{\partial_+\phi\partial_-\phi})+\frac12\partial_+\varphi\partial_-\varphi\right\} \,, \label{action.Abelian.example.flat.1}}
which is in fact a collection of three free bosons,  in accordance with \eqref{Eq:action.Abelian.zero-1}.
Another way to implement the $\lambda\to 0$ limit is to expand $\theta$ around $0$ (so that $\theta\to\sqrt{\lambda}\theta+\ldots$) and not to rescale $\phi$.\footnote{Rescaling also $\phi$, in this case, would make it disappear in the $\lambda\to 0$ limit. The need to keep $\phi$ finite in the $\lambda\to 0$ limit can be understood as implementing the scaling~\eqref{eq:lambdaZeroLimit-2} with $U$ depending on $\phi$. In fact, the $B$-field that appears above is exact, in agreement with~\eqref{Eq:action.Abelian.zero-2}.} This results in a $\sigma$-model on a flat target space and exact $B$-field, described by the action
\EQ{S=\frac{k}{2\pi}\int d^2x\left\{\alpha(\partial_+\theta\partial_-\theta+\theta^2\partial_+\phi\partial_-\phi)+\frac12\partial_+\varphi\partial_-\varphi-(\partial_+\phi\partial_-\varphi-\partial_+\varphi\partial_-\phi)\right\} \, , \label{action.Abelian.example.flat}}
which is in agreement with \eqref{Eq:action.Abelian.zero-2} and again equivalent just to a collection of three free bosons. 
The $\lambda \to 1$ limit requires a zoom-in as dictated in \eqref{limit.lambda.one}. In particular, we consider
\EQ{\lambda\to1-\frac{\kappa^2}{k}+\ldots\,,\quad \varphi\to\frac{\kappa^2}{k}\varphi+\ldots
\label{limit.lambda.one.ex}}
and send $k\to\infty$. The resulting action obtained from~\eqref{action.Abelian.example} reads
\EQ{\begin{split}
S=\frac{\kappa^2}{2\pi}\int d^2x&\left\{\alpha\left(\partial_+\theta\partial_-\theta+\sin^2\theta\partial_+\phi\partial_-\phi\right)+\partial_+\varphi\partial_-\varphi\right.\\
&\left.-\cos\theta(\partial_+\phi\partial_-\varphi-\partial_+\varphi\partial_-\phi)\right\}\,.
\end{split}\label{action.Abelian.one}}
As in \eqref{action.Abelian.example} this is the abelian T-dual of the {squashed} $S^3$ fibered over the $S^2$, {and for $\alpha=1$ it is of the round $S^3$.}

Let us now discuss the renormalisation of the action \eqref{action.Abelian.example} at one-loop order, that is at the leading order in the $k^{-1}$ expansion. We express the action as a non-linear $\sigma$-model on a target-space with metric $G_{\mu\nu}$ and antisymmetric field $B_{\mu\nu}$ (in the normalisation of footnote \ref{normalization.action})
\EQ{S=\frac{1}{2\pi}\int d^2x (G_{\mu\nu}+B_{\mu\nu})\partial_+X^\mu\partial_- X^\nu\,,\quad X^\mu=(\theta,\phi,\varphi)\, .}
Then, the one-loop renormalisation group (RG) flows are given in terms of \cite{Ecker:1971xko,Honerkamp:1971sh,Friedan:1980jf,Friedan:1980jm,Curtright:1984dz,Braaten:1985is,Fridling:1985hc} 
\EQ{\frac{d}{d\ln\mu^2} ( G_{\mu\nu} + B_{\mu\nu}) = \beta^{(1)}_{\mu\nu}+\nabla_\nu\xi_\mu+\frac12H_{\mu\nu\rho}\xi^\rho+\partial_\mu\zeta_\nu-\partial_\nu\zeta_\mu\,,\quad
\beta^{(1)}_{\mu\nu}=R^-_{\mu\nu}\,.
\label{RGs.one-loop}}
In the above, $\mu$ is the energy scale and $R^-_{\mu\nu}$ is the torsionful Ricci tensor given by
\EQ{
R^-_{\mu\nu}=R^{-\rho}{}_{\mu\rho\nu}=R_{\mu\nu}-\frac12\nabla^\rho H_{\mu\nu\rho}
-\frac14H_{\mu\kappa\lambda}H_\nu{}^{\kappa\lambda}\,,}
where $R^-_{\mu\nu\rho\lambda}$ is the torsionful Riemann tensor
\EQ{R^-_{\mu\nu\rho\lambda}=
R_{\mu\nu\rho\lambda}-
\frac12\left(\nabla_{\nu}H_{\mu\rho\lambda}
-\nabla_{\mu}H_{\nu\rho\lambda}\right)
+\frac14\left( H_{\rho\mu}{}^\kappa H_{\nu\lambda\kappa}-H_{\rho\nu}{}^\kappa H_{\mu\lambda\kappa}\right)\,.\label{generalized.Riemann}}
Moreover, $H_{\mu\nu\rho}$ denotes the components of the field strength associated with the $B$-field, and the covariant derivative is built using the Christoffel symbols of the metric $G_{\mu\nu}$. The vector $\xi=\xi^\mu\partial_\mu$ corresponds to field redefinitions (diffeomorphisms) $X^\mu\to X^\mu-\ln\mu^2\,\xi^\mu$,  and $\zeta=\zeta_\mu dx^\mu$ to gauge transformations of the two-form $B\to B-\ln\mu^2\,d\zeta$. The RG flow equations \eqref{RGs.one-loop} for the model \eqref{action.Abelian.example} translate into the following one-loop equations for $\lambda$ and $\alpha$ 
\EQ{\label{result.RGs}\beta^\lambda:=\frac{d\lambda}{d\ln\mu^2}=-\frac{\lambda^2}{2k\alpha^2}\,,\quad 
\beta^\alpha:=\frac{d\alpha}{d\ln\mu^2}=-\frac{\lambda}{2k\alpha }\frac{1-2\alpha+\lambda(3-2\alpha)}{1-\lambda^2}\,,}
where $k$ does not flow and the diffeomorphisms as well as the gauge transformations can be taken to be trivial. 
Let us now comment on some properties of the RG flows expressed by~\eqref{result.RGs}.
First of all, as a consequence of~\eqref{eq:sym-alpha0} and~\eqref{eq:sym-alpha}, the RG equations are invariant under the $\mathbb{Z}_2$ symmetry
\EQ{\lambda\to\lambda^{-1}\,,\quad \alpha\to\lambda^{-1}\alpha\,,\quad
k\to-k\,.\label{RG.symmetry}}
Moreover, it is possible to identify a combination of the couplings $\lambda,\alpha$ that is constant along the RG flow, namely
\EQ{\gamma^{(1)}:= \frac{\lambda(1+\lambda-2\alpha)}{\alpha(1-\lambda)^2}=\text{constant}\,.\label{eq:gamma}}
Of course, this combination is also invariant under the symmetry \eqref{RG.symmetry}. To study the RG flows \eqref{result.RGs},  we first solve \eqref{eq:gamma} in terms of $\alpha$ yielding\footnote{We thank Konstantinos Sfetsos for raising this up.} 
\EQ{\alpha=\frac{\lambda  (1+\lambda)}{\gamma^{(1)}  (1-\lambda )^2+2 \lambda }\,,
\label{alpha.oneloop}}
and upon inserting it into \eqref{result.RGs} we find a single RG for $\lambda$
\EQ{\beta^\lambda=-\frac{\left(2\lambda+\gamma^{(1)}  (1-\lambda)^2\right)^2}{2 k (1+\lambda)^2}\leqslant0\label{result.RGs.lambbda}\,,}
along with $\beta^{\gamma^{(1)}}=0$, resulting from \eqref{eq:gamma}.
Hence, $\lambda=0$ is not a UV fixed point of the RG flow for non-vanishing values of the constant $\gamma^{(1)}$. Note that the parameter $\alpha$ could depend on $\lambda$, such that $\alpha$ vanishes when $\lambda$ goes to zero with $\nicefrac{\lambda}{\alpha}$ being finite. Such a case is the abelian T-dual of the PCM on the round $S^3$ where $\alpha=\frac{2\lambda}{1+\lambda}$ (see footnote~\ref{foot.round}), yielding from \eqref{result.RGs} the RG flow
\begin{equation}
\beta^\lambda=-\frac{1}{8k}(1+\lambda)^2
\label{RG.S3.round}
\end{equation}
and $\lambda=0$ is clearly not a fixed point of the RG flow.

In general the UV fixed points of \eqref{result.RGs.lambbda} $\lambda_\pm$'s are given by
\EQ{\label{nontrivial.UV}
\lambda_+:=\frac{1}{\gamma^{(1)}}\left(\gamma^{(1)}-1+\sqrt{1-2\gamma^{(1)}}\right)\,,\quad \lambda_-=\lambda^{-1}_+\,,
}
provided $\gamma^{(1)}<\nicefrac12$, while demanding $\lambda\in(0,1)$ further restricts it to $\gamma^{(1)}<0$. We  focus on $\lambda_+$ that vanishes for $\gamma^{(1)}=0$, where the value of $\alpha$, as read from \eqref{alpha.oneloop}, simplifies to
\EQ{\alpha=\frac12(1+\lambda)\label{Eq:value.alpha}}and \eqref{result.RGs.lambbda} simplifies to the one-loop expression of the $\lambda$-deformed $SU(2)$ \cite{Itsios:2014lca} for $c_G=4$
\EQ{\beta^\lambda=-\frac{2\lambda^2}{k(1+\lambda)^2}\,,\label{beta.function.lambda}}
where $\lambda=0$ corresponds to a UV fixed point.
The choice of $\alpha$ in \eqref{Eq:value.alpha} matches with the value that was fixed by integrability in section~\ref{sec:int} in the case of $F$  non-abelian. However, notice that in the case at hand the subgroup $F$ is abelian, and the model is integrable for generic values of $\alpha$.

We may also compute the $C$-function of the model using the Weyl anomaly coefficient \cite{Tseytlin:1987bz,Tseytlin:2006ak}
\EQ{C=3-3\left(R-\frac{1}{12}H^2\right)=3+3\frac{\lambda  (\lambda -2 \alpha  (\lambda +1))}{\alpha ^2 k \left(1-\lambda ^2\right)}\,,}
and check that it is symmetric under \eqref{RG.symmetry} and that it satisfies Zamolodchikov's C-theorem \cite{Zamolodchikov:1986gt}
\EQ{\frac{dC}{d\ln\mu^2}=\frac{3 \lambda ^2 \left(2 \alpha ^2 (1+\lambda)^2-4 \alpha  (1+\lambda) \lambda +3 \lambda ^2\right)}{\alpha ^4 k^2 \left(1-\lambda ^2\right)^2}\geqslant0\,,}
where we have used \eqref{result.RGs}. 
Specialising to the choice of $\alpha$ in \eqref{Eq:value.alpha}, we find
\EQ{C=3-\frac{12 \lambda  \left(1+\lambda +\lambda^2\right)}{k (1-\lambda) (1+\lambda)^3}\,,\quad
\frac{dC}{d\ln\mu^2}=\frac{24\lambda^2 \left(1+4 \lambda ^2+\lambda ^4\right)}{k^2 (1-\lambda)^2 (1+\lambda)^6}\geqslant0\label{C-function.lambda}\,,}
which are distinct from those of the $\lambda$-deformed model found in \cite{Georgiou:2018vbb} despite the matching of their RG flows. The UV fixed points are distinct since   $\lambda=0$ in~\cite{Georgiou:2018vbb} corresponds to the WZW model of the $SU(2)_k$ whereas in the case at hand for $\lambda\to0$ we find free decoupled bosons, see Eq.\eqref{action.Abelian.example.flat.1}.
We also notice that the RG flow in~\eqref{result.RGs}  has a UV fixed point at $\lambda=0$ that, as noted above in~\eqref{action.Abelian.example.flat.1}, is just a collection of three free bosons.
Towards the IR, instead, if we take $\lambda\to1^-$ as in \eqref{limit.lambda.one.ex}, the RG flows in~\eqref{result.RGs} simplify to 
\EQ{\beta^\kappa:=\frac{d\kappa}{d\ln\mu^2}=\frac{1}{4\alpha^2\kappa}\,,\quad
\beta^\alpha:=\frac{d\alpha}{d\ln\mu^2}=\frac{\alpha-1}{\alpha\kappa^2}\,,}
As a consistency check we have derived these expressions starting from the action \eqref{action.Abelian.one} and using \eqref{RGs.one-loop}, where the diffeomorphisms and gauge transformations are identified by
\EQ{\xi=\frac{\varphi}{\alpha^2\kappa^2}\partial_\varphi\,,\quad \zeta=\frac{\varphi}{2\alpha^2}\cos\theta d\phi\,.}

We may actually study the RG flows of the $\sigma$-model \eqref{action.Abelian.example} to order $k^{-2}$ using the two-loop expressions \cite{Curtright:1984dz,Braaten:1985is,Hull:1987pc,Hull:1987yi,Metsaev:1987bc,Osborn:1989bu} 
 \EQ{&\frac{d}{d\ln\mu^2} ( G_{\mu\nu} + B_{\mu\nu}) = \beta^{(1)}_{\mu\nu}+\beta^{(2)}_{\mu\nu}
+\nabla_\nu\xi_\mu+\frac12H_{\mu\nu\rho}\xi^\rho+\partial_\mu\zeta_\nu-\partial_\nu\zeta_\mu\,,\\
&\beta^{(1)}_{\mu\nu}=R^-_{\mu\nu}\,,\quad
\beta^{(2)}_{\mu\nu}=
R^-_{\mu\kappa\lambda\rho}\left(R^{-\kappa\lambda\rho}{}_\nu-\frac12R^{-\lambda\rho\kappa}{}_\nu\right)+\frac12 (H^2)^{\lambda\rho}R^-_{\lambda\mu\nu\rho}\,,
\label{RGs.two-loop}}
where $(H^2)^{\lambda\rho}=H^{\lambda\kappa\tau}H^\rho{}_{\kappa\tau}$ and the generalised Riemann tensor was given in \eqref{generalized.Riemann}. Then we find
\AL{
\label{result.RGs.two}
&\beta^\lambda=-\frac{\lambda ^2}{2 k\alpha ^2 }+\frac{3 \lambda ^4}{2 \alpha ^4 \left(1-\lambda ^2\right) k^2}\,,\\
&\beta^\alpha=-\frac{\lambda}{2k\alpha}\frac{1-2\alpha+\lambda(3-2\alpha)}{1-\lambda^2}+
\frac{\lambda ^2 \left(4 \alpha ^2 (1+\lambda)^2-8 \alpha  \lambda  (1+\lambda)+3 \lambda  (1+3 \lambda)\right)}{2k^2\alpha ^3 \left(1-\lambda ^2\right)^2}\,,\nonumber
}
where again $k$ does not flow, and with trivial diffeomorphisms and gauge transformations. 
The system of the 2-loop RGs is invariant under the symmetry \eqref{RG.symmetry},\footnote{In contrast to the $\lambda$-deformed model, the symmetry in $k$ is not shifted by the quadratic Casimir of the group $G$ \cite{Kutasov:1989aw,Georgiou:2019nbz,Hoare:2019mcc}.} and (up to higher orders in the $k^{-1}$ expansion) it  admits the following constant along the RG flow\footnote{Upon differentiating with respect to the energy scale and using \eqref{result.RGs.two} we find that \eqref{integral.twoloop} is a constant to order $\nicefrac{1}{k^2}$.} 
\EQ{\gamma^{(1)}\left(1+k^{-1}\,f\left(\gamma^{(1)}\right)\right)-k^{-1}\frac{2 \lambda ^2 \left(2 \alpha ^2+(1-2 \alpha)  (1+\lambda)\right)}{ \alpha ^2 (1-\lambda )^4}=\text{constant}\,,\label{integral.twoloop}}
where $f(\gamma^{(1)})$ is an arbitrary function of $\gamma^{(1)}$ defined in~\eqref{eq:gamma}. Demanding invariance of the above expression under the symmetry \eqref{RG.symmetry} fixes $f(\gamma^{(1)})=\gamma^{(1)}$, and the constant \eqref{integral.twoloop} simplifies to
\EQ{\gamma^{(2)}:=\gamma^{(1)}-k^{-1}\frac{\lambda ^2 (1+\lambda)}{\alpha ^2 (1-\lambda)^3 }=\text{constant}\,.\label{eq:gamma.two}}
To study the two-loop RG flows \eqref{result.RGs.two} as in the one-loop order
we first solve \eqref{eq:gamma.two} in terms of $\alpha$ yielding at order $\nicefrac1k$,\footnote{The value of $\alpha=\frac{2\lambda}{1+\lambda}$ is not a consistent truncation of the two-loop expressions \eqref{result.RGs.two}, in contrast to the one-loop ones~\eqref{RG.S3.round}.}
\EQ{\alpha=\frac{\lambda  (1+\lambda)}{\gamma^{(2)}  (1-\lambda )^2+2 \lambda }-k^{-1}\frac{\lambda }{1-\lambda }
\label{alpha.twoloop}}
and upon inserting this value into \eqref{result.RGs.two} we find a single RG flow for $\lambda$
\EQ{
\beta^\lambda&=
-\frac{\left(2\lambda+\gamma^{(2)}  (1-\lambda)^2\right)^2}{2 k (1+\lambda)^2}\\
&-\frac{\left(2(1-\lambda+\lambda^2)-
3 \gamma^{(2)}  (1-\lambda)^2\right) \left(2 \lambda +\gamma^{(2)}  (1-\lambda)^2\right)^3}{2 k^2 (1-\lambda) (1+\lambda)^5}\,,\label{beta.function.lambda.two}
}
hence for a non-vanishing $\gamma^{(2)}$, the UV fixed point of the RG flow is given by \eqref{nontrivial.UV}, rather than $\lambda=0$ for $\gamma^{(2)}=0$. If $\gamma^{(2)}$ vanishes the value of $\alpha$ in
\eqref{alpha.twoloop} simplifies to
\EQ{\alpha=\frac12(1+\lambda)-k^{-1}\frac{\lambda}{1-\lambda}}
and the corresponding RG flow read from \eqref{beta.function.lambda.two} matches the two-loop expression for the $\lambda$-deformed $SU(2)$  \cite{Georgiou:2019nbz,Hoare:2019mcc} for $c_G=4$
\EQ{\beta^\lambda=-\frac{2\lambda^2}{k(1+\lambda)^2}-
\frac{8 \lambda ^3 \left(1-\lambda+\lambda^2\right)}{k^2(1-\lambda) (1+\lambda)^5 }\,.}
The $\lambda=0$ limit with the $\sigma$-model  \eqref{action.Abelian.example.flat.1}  corresponds again to a UV fixed point.
Taking the limit $\lambda\to1$ as in \eqref{limit.lambda.one.ex}, the RG flows \eqref{result.RGs.two} truncate to 
\EQ{\beta^\kappa=\frac{1}{4\alpha^2\kappa}-\frac{3}{8 \alpha ^4 \kappa ^3}\,,\quad
\beta^\alpha=\frac{\alpha-1}{\alpha\kappa^2}+\frac{3-4 \alpha+4 \alpha^2}{2\alpha ^3 \kappa ^4}\,.}
They can be derived from the action \eqref{action.Abelian.one} upon using the two-loop expressions \eqref{RGs.two-loop}, where the corresponding diffeomorphisms and gauge transformations are identified by
\EQ{\xi=\frac{\varphi}{\alpha^2\kappa^2}\left(1-\frac{3}{2\alpha^2\kappa^2}\right)\partial_\varphi\,,\quad \zeta=\frac{\varphi}{2\alpha^2}\left(1-\frac{3}{2\alpha^2\kappa^2}\right)\cos\theta d\phi\,.}

\section{Conclusions}\label{sec:concl}

We have constructed a new deformation of the PCM on a Lie group $G$, in such a way that only a subgroup $F\subset G$ is deformed, while the degrees of freedom in the coset  space $F\backslash G$ remain as spectators. 
The deformation is controlled by a parameter $\lambda\in (0,1)$, and the two limits $\lambda\to0$ and $\lambda\to1$ are correlated with the behaviour of the fields, to ensure that the deformed theory is described in terms of  
$\text{dim}(G)$ degrees of freedom for any value of $\lambda$.

Our construction is a generalisation of the $\lambda$-deformation first constructed in~\cite{Sfetsos:2013wia} (see also~\cite{Hollowood:2014rla,Hollowood:2014qma}), which is recovered when $F=G$.
We have assumed that the subgroup $F\subset G$ is chosen such that the restriction of the non-degenerate bilinear form of $G$ to $F$ is also non-degenerate. 
In particular, this ensures that $\mathfrak{f}^\ast= \mathfrak{f}$, where $\mathfrak{f}^\ast$ is the dual of $\mathfrak{f}$. Then, the Lie algebra of $G$ admits the orthogonal decomposition $\mathfrak{g}= \mathfrak{f}\oplus \mathfrak{f}^\perp$ such that $\Tr\{\mathfrak{f}\, \mathfrak{f}^\perp\}=0$.

Our main result is that the deformation preserves integrability if the Lie algebra structure satisfies the additional condition
\EQ{
\left[\mathfrak{f}^\perp, \mathfrak{f}^\perp\right]\subset \mathfrak{f},
}
which means that $F\backslash G$ is a symmetric space.
Then, the deformed theory interpolates between two points: at $\lambda=0$ we have the WZW action for $F$ together with a decoupled free boson that takes values in $\mathfrak{f}^\perp$, and at $\lambda=1$ the NATD action where only the subgroup $F$ is dualised.

Classical integrability is established by showing that the equations of motion of the deformed model can be written in terms of a Lax connection $\AA_\pm(z)$ with complex spectral parameter $z$. Remarkably, this Lax connection exhibits a square root branch cut, which means that we can make sense of it in terms of a 2-cover of the complex $z$-plane. We introduce an alternative choice of the spectral parameter such that the Lax connection becomes a meromorphic function with four single poles. Then, the deformation interpolates between $\lambda=0$, where the Lax connection vanishes, and $\lambda=1$, where it has two simple poles. For other values of $\lambda$ it always exhibits four poles. 

The requirements imposed by integrability are  very interesting. We may, in fact, compare our construction to other integrable deformations that are obtained by a similar strategy: first gauge a symmetry of the model, then add a topological term to the action that is compatible with the gauge symmetry, and finally integrate out some fields. See for example~\cite{Sfetsos:2013wia,Hollowood:2014rla} where the topological term is a topological gauged WZW action, and~\cite{Hoare:2016wsk,Borsato:2016pas} where the topological term is constructed from a 2-cocycle on the Lie algebra. In those cases, on top of producing  explicit Lax connections, the integrability of the deformations was also argued to be a natural consequence of the addition of the \emph{topological} term, that therefore does not modify the equations of motion of the original integrable model.  It may be tempting to expect that a similar argument should apply also in the case of the $\lambda$-deformations constructed here. However, as already noted, we find that integrability is not preserved in general, but only when $F\backslash G$ is symmetric and $\alpha=\frac{1+\lambda}{2}$. These features are, in fact, reminiscent of similar observations made in the context of the $\lambda$-deformation of the ${AdS}_5\times {S}^5$ superstring. In fact, in~\cite{Hollowood:2014qma} it was observed that just adding a gauged WZW term to the action fixes the form of the integrable deformation only in the leading order in $\nicefrac1k$. In general, one should not expect that coupling two integrable models through a gauging procedure should necessarily preserve integrability. One obvious example are coset $\sigma$-models, that may be obtained by gauging symmetries of a PCM but are not always integrable, unless the gauging respects a $\mathbb Z_2$ structure~\cite{Eichenherr:1979ci} or more general $\mathbb Z_n$ structure~\cite{Young:2005jv}.
It would be nice to understand why  the argument does work in certain cases, for example in the case of~\cite{Sfetsos:2013wia}.\footnote{We thank B. Hoare for discussions on these points.}

It would be interesting to understand how to generalise our construction by relaxing the nondegeneracy requirement of the restriction to $\mathfrak{f}$ of the bilinear form on $\mathfrak{g}$. As explained in section~\ref{sec:deg}, the main obstacle to do so is related to the impossibility to write a WZW model on $F$ if the corresponding bilinear form is degenerate. This problem was addressed by Nappi and Witten for the  WZW model in~\cite{Nappi:1993ie}, although in our case we have to consider the restriction of a non-degenerate bilinear form on $G$ to a subgroup $F$. The usual $\lambda$-deformation of the Nappi--Witten model was recently considered in \cite{Sfetsos:2022irp}.

One may also try to write this construction in the language of 4D and 6D Chern--Simons theories, which was recently used to reformulate the original $\lambda$-deformation and other integrable deformations of $\sigma$-models~\cite{Delduc:2019whp,Cole:2023umd,Cole:2024sje} following the approach of~\cite{Costello:2017dso,Costello:2018gyb,Costello:2019tri}.

Other aspects of these theories, which are left for future projects, include their Hamiltonian description and its possible formulation in terms of $q$-deformed algebras. It is well known that the Poisson brackets of the original $\lambda$ deformation can be described in terms of a $q$ deformed algebra with $q$ being a root of unity~\cite{Hollowood:2014rla}.\footnote{In~\cite{Hollowood:2014rla}, $k$ is assumed to be an integer and, thus, $q=e^{i\nicefrac{\pi}{k}}$.} It would be interesting to understand if our models implement a partial $q$-deformation of the Lie algebra of $G$, where only $\mathfrak{f}$ is deformed. In particular, it would be interesting to understand if the $\mathbb Z_2$ symmetry decomposition is a requirement that emerges when constructing a $q$-deformation of $F\subset G$.

Another extension of the present work involves the study of the one-loop renormalisation group flows for the parameters $\lambda$ and $\alpha$ in \eqref{eq:S-lambda-sub}, extending the results found in the example of section \ref{sec:ex}
to generic semisimple group $G$ and subgroup $F$. This may be done by calculating the one-loop effective action with heat kernel techniques as in~\cite{Appadu:2015nfa,Sagkrioti:2018rwg,Delduc:2020vxy}. We expect that at least for the integrable sector studied in \ref{sec:int}, where in most cases $\alpha=\frac12(1+\lambda)$,
it should be renormalisable as argued in \cite{Hoare:2020fye,Delduc:2020vxy,Levine:2022hpv}. 

Finally, it would be interesting to explore a possible embedding of our construction to solutions of type-II supergravity with non-trivial Ramond–-Ramond fields. For the $\lambda$-deformed model of~\cite{Sfetsos:2013wia}, related embeddings have been constructed for deformations of groups and symmetric cosets~\cite{Sfetsos:2014cea,Demulder:2015lva,Itsios:2019izt,Itsios:2023kma,Itsios:2023uae}, and the same brute-force strategy may be attempted in our case.   
Another natural strategy to achieve this is to generalise our construction first to the case of symmetric cosets and then semisymmetric supercosets. These are known to admit a Lax connection and be classically integrable, and in some cases they also give rise to superstring backgrounds with non-trivial Ramond--Ramond fluxes, like the $AdS_5\times S^5$ one. It would be very interesting to understand the options to construct a $\lambda$-deformation of these models when only a subgroup of the global symmetries is deformed. One motivation for these studies is to explore the possibilities to construct new $\lambda$-deformations of backgrounds that appear in instances of the AdS/CFT correspondence.


\section*{Acknowledgements}
We thank Ben Hoare, Tim Hollowood and Konstantinos Sfetsos for discussions, as well as Ben Hoare and Konstantinos Sfetsos for comments on the draft.
The work of R.~B. and J.L.~M. was supported by AEI-Spain (under project PID2020-114157GB-I00 and Unidad de Excelencia Mar\'\i a de Maetzu MDM-2016-0692),  Xunta de Galicia (Centro singular de investigaci\'on de Galicia accreditation 2023-2026, and project ED431C-2021/14), and the European Union FEDER.  R.~B. is also supported by the grant RYC2021-032371-I (funded by MCIN/AEI/10.13039/501100011033 and by the European Union ``NextGenerationEU''/PRTR) and by the grant 2023-PG083 with reference code ED431F 2023/19 funded by Xunta de Galicia.
The research work of G.I.  is supported by the Einstein Stiftung Berlin via the Einstein International Postdoctoral Fellowship program 
 ``Generalised dualities and their holographic applications to condensed matter physics'' (project number IPF-2020-604). G.~I. is also supported by the Deutsche Forschungsgemeinschaft (DFG, German Research Foundation) via the Emmy Noether program ``Exploring the landscape of string theory flux vacua using exceptional field theory'' (project number 426510644).

\appendix

\section{Derivation of the equations of motion}
\label{app:equations-of-motion}

The transformation of the two components of~\eqref{eq:Action} under infinitesimal variations of the fields is
\AL{
\delta \widetilde S_\text{gPCM}[g,A_\mu;\alpha]=&
 -\frac{\kappa^2}{\pi}\int d^2 x\, \Tr\left\{-\delta g g^{-1}\left(\left[\partial_+ +A_+, j_-^f+A_-\right]
 +\left[\partial_- +A_-, j_+^f+A_+\right]\right. \right.\nonumber\\[5pt]
 &\qquad
 \left.\left. +\left[\alpha\partial_+ + (1-\alpha)j_+^f + A_+,j_-^\perp\right]+ 
 \left[\alpha\partial_- + (1-\alpha)j_-^f + A_-,j_+^\perp\right]\right)\right\}\nonumber\\[5pt]
&
-\frac{\kappa^2}{\pi}\int d^2 x\, \Tr\left\{
\delta A_+ \left(j_-^f + A_-\right) + \delta A_- \left(j_+^f + A_+\right)\right\},
}
and
\EQ{
&
\delta S_\text{gWZW}[{\cal F},A_\mu]
=-\frac{k}{\pi} \int d^2x\, \Tr\left\{\CF^{-1}\delta\CF\left[\partial_+ +\CF^{-1}\partial_+\CF+\CF^{-1}A_+\CF,\partial_-+A_-\right]\right.\\[5pt]
&
\quad
\left. -\delta A_+ \left(-\partial_-\CF\CF^{-1} +\CF A_-\CF^{-1}- A_-\right) - \delta A_- \left(\CF^{-1}\partial_+\CF +\CF^{-1} A_+\CF- A_+\right)\right\},
\label{TransWZW}
}
where
\EQ{
j_\pm = \partial_\pm g g^{-1} = j^\mathfrak{f}_\pm + j^\perp_\pm.
}
Then, the equations of motion of the gWZW field $\CF$ read
\EQ{
\left[\partial_+ +\CF^{-1}\partial_+\CF+\CF^{-1}A_+\CF,\partial_-+A_-\right]=0\,,
\label{eomF1}
}
which, by conjugating with~$\CF$, are equivalent to
\EQ{
\left[\partial_+ + A_+,\partial_- -\partial_-\CF \CF^{-1} +\CF A_-\CF^{-1}\right]=0.
\label{eomF2}
}
Using the decomposition~\eqref{eq:decomp-f-fperp}, $\mathfrak{g}=\mathfrak{f}\oplus \mathfrak{f}^\perp$, the equations of motion of the PCM field $g$ become
\AL{
&
\left[\partial_+ +A_+, j_-^f+A_-\right]
 +\left[\partial_- +A_-, j_+^f+A_+\right]=0,\label{eomf-f}\\[5pt]
 &
 \left[\alpha\partial_+ + (1-\alpha)j_+^f + A_+,j_-^\perp\right]+ 
 \left[\alpha\partial_- + (1-\alpha)j_-^f + A_-,j_+^\perp\right]=0,
 \label{eomf}
 }
together with the Maurer--Cartan identity
\EQ{
\left[\partial_+ - j_+, \partial_- - j_-\right]=0.
\label{eq:MCidentity-app}
}
Finally, the equations of motion of the gauge fields are
\EQ{
&
\kappa^2 \left(j_-^\mathfrak{f}+A_-\right)=k\left(-\partial_-\CF\CF^{-1} +\CF A_-\CF^{-1}-A_-\right),\\[5pt]
&
\kappa^2\left(j_+^\mathfrak{f}+A_+\right)=k\left(\CF^{-1}\partial_+\CF +\CF^{-1} A_+\CF-A_+\right)
\label{eomA}
}
and, as  usually done for this kind of deformation, we  parameterise the relative magnitude of the two couplings $\kappa$ and $k$ with the parameter
\EQ{
\lambda=\frac{k}{k+\kappa^2}\qquad \Longrightarrow \qquad \frac{\kappa^2}{k}= \frac{1-\lambda}{\lambda}.
}

The identities~\eqref{eomA} allow one to write the equations of motion of the gWZW field $\CF$ in terms of the PCM fields $g$ and $A_\pm$. Namely, they allow one to write
\SP{
&
-\partial_-\CF \CF^{-1}+ \CF A_-\CF^{-1} = \frac{1-\lambda}{\lambda} \, j_-^\mathfrak{f} + \frac{1}{\lambda}\, A_-,\\[5pt]
&
\CF^{-1} \partial_+\CF + \CF^{-1} A_+\CF = \frac{1-\lambda}{\lambda} \, j_+^\mathfrak{f} + \frac{1}{\lambda}\, A_+
}
and, thus, eqs.~\eqref{eomF1} and~\eqref{eomF2} become
\EQ{
\left[\partial_\pm + \frac{1-\lambda}{\lambda} j_\pm^\mathfrak{f} + \frac{1}{\lambda} A_\pm, \partial_\mp + A_\mp\right]=0.
\label{eomF-new}
}
Using~\eqref{eomf-f}, these two equations are identical.

Therefore, the set of independent equations of motion is provided by~\eqref{eomF-new}, \eqref{eomf} and the Maurer--Cartan identity~\eqref{eq:MCidentity-app}, which are those used in section~\ref{sec:int} (eqs.~\eqref{eq:eomF}-\eqref{eq:MCidentity}). In addition, the identities~\eqref{eomA} allow one to write the fields $A_\pm$ in terms of $j_\pm^\mathfrak{f}$ and $\CF$, which would provide an equivalent way to present those equations in terms of the gauged WZW field $\CF$ and the PCM currents $j_\pm=\partial_\pm g g^{-1}$. All these equations are invariant under the gauge transformations~\eqref{eq:GaugeTranf}.

\bibliographystyle{nb}
\bibliography{biblio}{}

\end{document}